# Mathematical model of subthalamic nucleus neuron: characteristic activity patterns and bifurcation analysis

Choongseok Park[1*], Leonid L. Rubchinsky[2,3], Sungwoo Ahn[4]

[1]Department of Mathematics and Statistics, North Carolina A&T State University, Greensboro, NC, USA 27411
[2] Department of Mathematical Sciences, Indiana University-Purdue University Indianapolis, Indianapolis, IN, USA 46202
[3] Stark Neurosciences Research Institute, Indiana University School of Medicine, Indianapolis, IN, USA 46202
[4]Department of Mathematics, East Carolina University, Greenville NC, USA 27858

* Correspondence:
Choongseok Park
Department of Mathematics and Statistics
North Carolina A&T State University
1601 E Market St
Greensboro, NC 27411, USA
Email: cpark@ncat.edu

## Abstract

The subthalamic nucleus (STN) has an important role in the pathophysiology of the basal ganglia in Parkinson's disease. The ability of STN cells to generate bursting rhythms under either transient or sustained hyperpolarization may underlie the excessively synchronous beta rhythms observed in Parkinson's disease. In this study, we developed a conductance-based single compartment model of an STN neuron, which is able to generate characteristic activity patterns observed in experiments including hyperpolarization-induced bursts and post-inhibitory rebound bursts. This study focused on the role of three currents in rhythm generation: T-type calcium (CaT) current, L-type calcium (CaL) current, and hyperpolarization-activated cyclic nucleotide-gated (HCN) current. To investigate the effects of these currents in rhythm generation, we performed a bifurcation analysis using slow variables in these currents. Bifurcation analysis showed that the HCN current promotes single-spike activity patterns rather than bursting in agreement with experimental results. It also showed that the CaT current is necessary for characteristic bursting activity patterns. In particular, the CaT current enables STN neurons to generate these activity patterns under hyperpolarizing stimuli. The CaL current enriches and reinforces these characteristic activity patterns. In hyperpolarization-induced bursts or post-inhibitory rebound bursts, the CaL current allows STN neurons to generate long bursting patterns. Thus, bifurcation analysis explained the synergistic interaction of the CaT and CaL currents, which enables STN neurons to respond to hyperpolarizing stimuli in a salient way. The results of this study implicate the importance of CaT and CaL currents in the pathophysiology of the basal ganglia in Parkinson's disease.

The subthalamic nucleus is an important part of the brain due to its involvement in motor behavior and other brain functions. Pathological activity in subthalamic neurons contributes to the motor symptoms of Parkinson's disease. Therefore, it is important to understand the mechanistic basis of electrical activity in subthalamic neurons. These neurons are known to exhibit several distinct patterns of activity: tonic spikes (action potentials), hyperpolarization-induced bursts of spikes, and post-inhibitory rebound bursts of spikes. This study explores how different membrane channels interact with each other to give rise to various activity patterns in a single cell within the framework of a one-compartment (i.e. no spatial extent) model. Bifurcation analysis using slow variables in membrane currents showed how calcium currents can promote different types of bursting, especially in response to an inhibitory input. The study also emphasizes the importance of slow calcium currents in subthalamic cells in Parkinson's disease known for its bursting electrical activity in the subthalamus.

# 1 Introduction

The basal ganglia (BG) is a group of interconnected subcortical nuclei, which is involved in the generation of movement, cognition, and emotion. It is known to be impacted in Parkinson's disease. Excessively synchronized beta-frequency band rhythms (13-30 Hz) within the BG have been reported in parkinsonian patients and in parkinsonian animal models [Brown, 2003; Hammond et al., 2007; Mallet et al., 2008; Park et al., 2010; Oswal et al., 2013; Stein and Bar-Gad, 2013; Ahn et al., 2015]. These rhythms have been implicated in the motor symptoms of Parkinson's Disease (PD), such as slowness and rigidity of movements [Hutchison et al., 2004; Kühn et al., 2004; Brown, 2007; Ray et al., 2008; Eusebio and Brown, 2009; Kühn et al., 2009]. As the only excitatory nucleus in the BG, the subthalamic nucleus (STN) plays an important role in the dynamics and functions of the BG [Bevan et al., 2002b; Kühn et al., 2009; Hirschmann et al., 2011; Tachibana et al., 2011; Pavlides at al., 2015; Ahn et al., 2016; Rubin, 2017]. The STN is also a standard target for electrical deep brain stimulation (DBS), a commonly used treatment for advanced PD patients [Wingeier et al., 2006; Kühn et al., 2008; Eusebio et al., 2011].

In vitro, the STN neuron fires single spikes in a slow rhythmic manner at 5-20 Hz during the absence of external input, which may underlie the tonic firing patterns observed in vivo in resting animals [Beurrier et al., 1999; Bevan and Wilson, 1999]. STN firing frequency increases almost linearly with the magnitude of injected depolarizing current pulses [Hallworth et al., 2003; Wilson et al., 2004]. Bevan et al. showed that a majority of STN neurons elicit a calcium-dependent post-inhibitory rebound (PIR) burst of spikes when the neurons are released from inhibitory synaptic input [Bevan et al., 2002a]. PIR bursts can be either long or short depending on the level and duration of inhibition received by the neuron. On the other hand, some STN neurons under hyperpolarized conditions switched from a spontaneously discharging single-spike mode to a pure burst-firing mode (consisting of long-lasting bursts of constant duration) or a mixed burst-firing mode (alternating short and long bursts) [Beurrier et al., 1999]. It was argued that slow rhythmic bursting results from T-type calcium currents and L-type calcium currents [Beurrier et al., 1999; Bevan and Wilson, 1999].

Since the STN receives inhibitory input from the external globus pallidus (GPe), these characteristic activity patterns may be essential ingredients in the normal and abnormal functioning of the system. For example, STN neurons are able to transform sustained inhibitory synaptic input into rhythmic bursts of spikes. Hence, their resting activities can transition from tonic spiking discharge to burst-firing patterns. Therefore, hyperpolarization-induced bursting in STN neurons may play crucial roles in the generation of excessively synchronized rhythmic bursting patterns observed in parkinsonian BG (see references above).

The goal of this study is to develop a relatively simple STN neuron model which exhibits several activity patterns as described above, and to understand the dynamic mechanisms of these patterns by studying the interactions of membrane currents and the bifurcation diagrams underlying transitions between different activity modes. A model, which is able to generate all types of activity, was developed earlier [Gillies and Willshaw, 2006], but it is a complicated multi-compartmental model where the interaction of compartments appears to be essential for its dynamical regimes. While this model is able to generate the activity patterns described above, mathematical analysis of the underlying mechanisms is very challenging due to the complexity of the multi-compartmental model. In addition, the model's complexity makes it hard to use in the construction of a large model network, such as the cortex-BG network to study the mathematical mechanisms of synchronized beta rhythms in networks.

In our current study, we developed a conductance-based single compartment model of the STN neuron and showed how this model captures the characteristic activity patterns, especially hyperpolarization-induced bursting and post-inhibitory rebound (PIR) bursting. Using this model, we performed a bifurcation analysis to study the roles and effects of three currents (T-type calcium current, L-type calcium current, and hyperpolarization-activated cyclic nucleotide-gated current) in the rhythm generation mechanisms under inhibition [Beurrier et al., 1999; Atherton et al., 2010].

## 2 Mathematical model

A conductance-based single compartment model of the STN neuron includes spike-generating potassium and sodium currents ($I_K$ and $I_{Na}$), a leak current ($I_L$), a persistent sodium current ($I_{NaP}$), a calcium dependent potassium current ($I_{AHP}$), hyperpolarization-activated cyclic nucleotide-gated (HCN) current ($I_{HCN}$), an A-type potassium current ($I_A$), a T-type low-threshold calcium (CaT) current ($I_{CaT}$), and an L-type high-threshold calcium (CaL) current ($I_{CaL}$). For basic currents (spike-generating currents, leak currents, and AHP currents), we used the equations in the Terman model [Terman et al., 2002; Best et al., 2007]. The forms of A-type calcium current, CaT current and CaL current were adopted from the Hahn and McIntyre model [Hahn and McIntyre, 2010]. The form of the HCN current was based on the Gillies and Willshaw model [Gillies and Willshaw, 2006]. The dynamics of the membrane potential ($V$) is described by the following differential equations:

$$C\frac{dV}{dt} = -I_L - I_K - I_{Na} - I_{NaP} - I_{AHP} - I_{HCN} - I_A - I_{CaT} - I_{CaL} - I_{app0} - I_{app} \quad (1)$$

$$\frac{dx}{dt} = \frac{x_\infty(V)-x}{\tau_x(V)}, \quad x \in \{m, h, n, f, a, b, p, q, c, d_1\} \quad (2)$$

$$\frac{dx}{dt} = \frac{x_\infty([Ca])-x}{\tau_x(V)}, \quad x \in \{r, d_2\} \quad (3)$$

$$\frac{d[Ca]}{dt} = \frac{\epsilon}{2F}(-I_T - I_{CaL}) - K_{Ca}[Ca] \quad (4)$$

where $I_L = g_L(V - V_L)$, $I_K = g_K n^4 (V - V_K)$, $I_{Na} = g_{Na} m^3 h (V - V_{Na})$, $I_{NaP} = g_{NaP}(V - V_{Na})$, $I_{AHP} = g_{AHP} r^2 (V - V_K)$, $I_{HCN} = g_{HCN} f (V - V_{HCN})$, $I_A = g_A a^2 b (V - V_K)$, $I_{CaT} = g_{CaT} p^2 q (V - V_{Ca})$, $I_{CaL} = g_{CaL} c^2 d_1 d_2 (V - V_{Ca})$. The units for ionic currents are mA/cm$^2$. In the first equation, $I_{app0}$ is the baseline external input, and $I_{app}$ represents an injected applied current. In the last equation, [Ca] is the calcium concentration in mM, $F$ is the Faraday's constant and $K_{Ca}$ = 0.2/ms is the calcium pump rate. Voltage dependent activation and inactivation steady states and time constants are given as follows:

$$x_\infty(V) = \left[1 + \exp\left(\frac{V - \theta_{\infty,x}}{\sigma_{\infty,x}}\right)\right]^{-1}, \quad x \in \{m, h, n, f, a, b, p, q, c, d_1\} \quad (5)$$

$$x_\infty([Ca]) = \left[1 + \exp\left(\frac{[Ca] - \theta_{\infty,x}}{\sigma_{\infty,x}}\right)\right]^{-1}, \quad x \in \{r, d_2\} \quad (6)$$

$$\tau_x(V) = \tau_{0,x} + \tau_{1,x}\left[1 + \exp\left(-\frac{V - \theta_{1,x}}{\sigma_{1,x}}\right)\right]^{-1} + \tau_{2,x} \exp\left(-\frac{V - \theta_{2,x}}{\sigma_{2,x}}\right), \quad (7)$$

$$x \in \{m, h, n, r, a, b, p, q, c, d_1, d_2\} \quad (8)$$

$$\tau_f(V) = \tau_{0,f} + \tau_{1,f}\left[\exp(\theta_{1,f} + \sigma_{1,f}V) + \exp(\theta_{2,f} + \sigma_{2,f}V)\right]^{-1} \quad (9)$$

The values of maximal conductances are as follows: $g_L = 0.9\ S/cm^2$, $g_K = 57\ S/cm^2$, $g_{Na} = 49\ S/cm^2$, $g_{NaP} = 0.003\ S/cm^2$, $g_{AHP} = 1\ S/cm^2$, $g_{HCN} = 2\ S/cm^2$, $g_A = 5\ S/cm^2$, $g_{CaT} = 20\ S/cm^2$, $g_{CaL} = 5\ S/cm^2$. The values of reversal potentials are as follows: $V_L = -60\ mV$, $V_K = -80\ mV$, $V_{Na} = 55\ mV$, $V_{HCN} = -43\ mV$, $V_{Ca} = 120\ mV$. Kinetic parameter values were obtained from the Hahn and McIntyre model [Hahn and McIntyre, 2010] and the Gillies and Willshaw model [Gillies and Willshaw, 2006], then tuned to capture the characteristic activity patterns, especially hyperpolarization-induced bursts and PIR bursts. Table 1 lists the resulting values of the kinetic parameters.

|  | $\theta_{\infty,x}$ (mV) | $\sigma_{\infty,x}$ (mV) | $\tau_{0,x}$ (msec) | $\tau_{1,x}$ (msec) | $\tau_{2,x}$ (msec) | $\theta_{1,x}$ (mV) | $\sigma_{1,x}$ (mV) | $\theta_{2,x}$ (mV) | $\sigma_{2,x}$ (mV) |
|---|---|---|---|---|---|---|---|---|---|
| m | -40 | -8 | 0.2 | 3 | 0 | -53 | -0.7 | | |
| h | -45.5 | 6.4 | 0 | 24.5 | 1 | -50 | -10 | -50 | 20 |
| n | -41.5 | -14 | 0 | 11 | 1 | -40 | -40 | -40 | 50 |
| r | 0.17 (mM) | -0.08 (mM) | 2 | 0 | 0 | | | | |
| f | -75 | 5.5 | 0 | 1 | 1 | -14.59 | -0.086 | -1.87 | 0.08 |
| a | -45 | -14.7 | 1 | 1 | 0 | -40 | -0.5 | | |
| b | -90 | 7.5 | 0 | 200 | 1 | -60 | -30 | -40 | 10 |
| p | -56 | -6.7 | 1 | 0.33 | 200 | -27 | -10 | -102 | 15 |
| q | -85 | 5.8 | 0 | 400 | 100 | -50 | -15 | -50 | 16 |
| c | -30.6 | -5 | 45 | 10 | 15 | -27 | -20 | -50 | 15 |
| $d_1$ | -60 | 7.5 | 400 | 500 | 1 | -40 | -15 | -20 | 20 |
| $d_2$ | 0.2 (mM) | 0.02 (mM) | 3000 | 0 | 0 | | | | |

**Table 1.** Values of kinetic parameters. Units for each parameter values are shown in the first row except $\theta_{\infty,r}$, $\theta_{\infty,d2}$, $\sigma_{\infty,r}$, $\sigma_{\infty,d2}$ whose units are mM.

In the following sections, we will focus on CaT, CaL, and HCN currents and study the effects and interactions of these currents on hyperpolarization-induced rhythms. In the next section, we begin with spontaneous tonic firing activity. Although spontaneous tonic spiking is not induced by hyperpolarization, the results in that section will be used to explain the effect and interaction of CaT and CaL currents on hyperpolarization-induced rhythms. Note that in the rest of the manuscript, we will omit the unit of each parameter for the sake of brevity and simplicity. Bifurcation parameters in the following sections are formed by multiplying maximal conductances and dimensionless gating variables, thus they follow the same units as the maximal conductances used therein.

### 3 Spontaneous tonic firing activity

The model exhibits spontaneous tonic firing activity (~10 Hz), which depends on sodium, potassium, and persistent sodium currents. Figure 1 shows the dependence of these rhythms on the following four intrinsic parameters: the magnitude of the injected depolarization current ($I_{app}$), the maximal conductance of the CaT current ($g_{CaT}$), the maximal conductance of the HCN current ($g_{HCN}$), and the maximal conductance of the CaL current ($g_{CaL}$). Default parameter values are $g_{CaT} = 20$, $g_{CaL} = 5$, $g_{HCN} = 2$, and $I_{app} = 0$.

As observed in experiments and discussed in other studies [Hallworth et al., 2003; Wilson et al., 2004], firing frequency increases almost linearly as the magnitude of injected depolarizing current increases (Fig. 1A) or as $g_{CaT}$ increases (Fig. 1B). There is only a slight change in frequency when $g_{HCN}$ increases (Fig. 1C). In the case of CaL current variation, however, the frequency shows an abrupt jump around $g_{CaL} = 17$ while the frequencies are almost constants for small or large values of $g_{CaL}$ (Fig. 1D). In Sec 3.2, we will explore how the CaL current and other currents interact to generate these firing patterns through a bifurcation analysis. Note that, when we increased the timescale of the HCN gating variable $f$ dynamics by 50% (that is, the value of $\tau_f$ was multiplied by 1.5. Note that $\tau_{0,f} = 0$.) in Fig. 1D, there is almost no change in frequency for a wide range of $g_{CaL}$ values. In Sec 3.2, we will discuss this observation in detail.

The mechanism underlying the effect of applied current on spontaneous tonic firing activity has been extensively studied in different types of neuron models [Terman, 1992; Ermentrout and Terman, 2010; Park and Rubin, 2013]. The response of these neuron models and the model considered here to the constant applied current are similar and have been considered in the studies mentioned above. Therefore, we will focus on the effect of the three currents considered here (CaT, CaL, and HCN currents) on spontaneous tonic firing activity. In the following sections, we study how the role and mechanisms of these currents interplay in each characteristic firing pattern using bifurcation analysis. We will utilize the separation of timescales technique and slow variables to construct bifurcation diagrams (the fast-slow analysis).

### 3.1 Effect of CaT and HCN currents on spontaneous tonic firing rhythm. Bifurcation analysis.

To study the effects of the CaT current and the HCN current on spontaneous tonic firing rhythms, we used a fast-slow analysis with $g_{CaL}$ fixed at 5. In the standard fast-slow analysis, we consider slow variables as bifurcation parameters and derive a bifurcation diagram of the fast subsystem. During spontaneous tonic firing activity in the model, the level of $[Ca]$ is very low and changes slowly, which is due to the slow timescale of $[Ca]$ and low spiking frequency. Hence, we may disregard the effect of $[Ca]$ and the CaL current in this case. Note that the gating variables of the CaT current and the HCN current do not depend on $[Ca]$. On the other hand, it is known that the HCN current promotes a single-spike activity [Atherton et al., 2010]. For this reason, we used the following slow variables $ghcnc = g_{HCN} * f$ and $gtc = g_{CaT} * q$ for the analysis, where $f$ and $q$ are the gating variables in HCN and CaT currents, respectively. In summary, $f$ and $q$ (or $ghcnc$ and $gtc$) are slow variables and the remaining variables are then fast variables. The system of governing equations of the fast variables forms a fast subsystem.

Figure 2A shows one example of a bifurcation diagram of the fast subsystem, which was projected onto the $(gtc, V)$-space. Here we treated $gtc$ as a bifurcation parameter and fixed $ghcnc$ at 0.01. The set of fixed points of the fast subsystem forms an S-shaped curve, $S$, in the $(gtc, V)$-space and this structure persists over a certain range of $ghcnc$ values of interest. The lower branch of $S$ at the lower left corner (black solid) consists of stable fixed points and the middle branch of $S$ unstable saddle points (black dashed). The lower branch and middle branch coalesce at a fold bifurcation, which we call the right knee (RK) of $S$. Similarly, this middle branch turns around at another fold bifurcation point. We call this upper fold bifurcation point the left knee (LK) of $S$. As $gtc$ increases from LK, the fast subsystem undergoes an Andronov-Hopf (AH) bifurcation, at a value of $gtc$ that we denote by $gtc_{AH}$, and above this, the fixed points become stable. A family of stable periodic orbits ($P$) emerges from $S$ at the AH-point. The two black curves from the AH-point show the minimum and maximum $V$ along the family of periodic orbits. Finally, the family of stable periodic orbits ($P$) terminates in a saddle-node on invariant circle (SNIC) bifurcation. Figure 2A also shows the projection of the spontaneous tonic firing solution (blue) of the full model onto the bifurcation diagram when $g_{CaT} = 20$, along with the corresponding $gtc$-nullcline (green). To the right (above) of the $gtc$-nullcline, $gtc' < 0$; hence, $gtc$ decreases over that region as a dynamic variable. Similarly, $gtc' > 0$ to the left (below) of the $gtc$-nullcline. Note that $gtc$-nullcline lies above the lower branch of $S$ in the diagram. As will be explained later in this section, two characteristics of the bifurcation diagram -- 1) the family of stable periodic orbits terminates at the SNIC and 2) the $gtc$-nullcline lies above the lower branch of $S$-- are crucial for the generation of tonic spiking solutions. For this reason, this type of bifurcation has been frequently associated with tonic spiking solutions. As will be explained in Section 4 (cf. Figures 4 and 5), if the stable periodic orbits terminate in a homoclinic orbit, then a square-wave bursting in the neural model is more likely to occur. [Rinzel, 1987; Best et al., 2005; Butera et al., 2005; Bertram and Rubin, 2017]

Now treating $gtc$ and $ghcnc$ as bifurcation parameters, a bifurcation diagram in Fig. 2A forms a bifurcation surface in the ($gtc$, $ghcnc$, $V$)-space (Fig. 2B). The black surfaces on the top and bottom of the right side of the figure are surfaces formed by maximum and minimum values of $V$ along the families of stable periodic orbits (cf. Fig. 2A). Let $\Sigma$ denote the surface formed by S-shaped curve $S$ from the Fig. 2A. Part of the surface $\Sigma$ is also shown between the surfaces of stable periodic orbits. More specifically, the black surface that crosses the figure horizontally in the middle is the surface of unstable fixed points. The folded surface corresponds to the lower and middle branches of $\Sigma$. This surface is folded at the line of RK points, separating stable and unstable points. Figure 2B also shows the projection of the spontaneous tonic firing solution when $g_{CaT}$ = 20 (blue), the same solution as used in Fig. 2A. The $gtc$-nullsurface is also shown in green. This $gtc$-nullsurface divides the phase space into two parts. In the region below the $gtc$-nullsurface, $gtc' > 0$, hence $gtc$ increases. On the other hand, in the region above the surface, $gtc' < 0$ and $gtc$ decreases.

The projection of tonic spiking solution sweeps both regions. Figure 2B shows that the projected trajectory jumps down to the lower part of $\Sigma$. Consider a projected trajectory that lies in a small neighborhood of the lower part of $\Sigma$ (surface of stable fixed points). Because the lower part of $\Sigma$ lies below the $gtc$-nullsurface ($gtc' > 0$), $gtc$ increases along the surface. In this case, $ghcnc$ also increases and as a result, the trajectory traverses the lower part of Σ. Once the trajectory reaches the line of RK points (the boundary between the lower and middle parts of Σ), then the trajectory jumps up into the regime of stable periodic orbits of fast subsystem. While staying in the regime of stable periodic orbits, both $gtc$ and $ghcnc$ decrease, hence the trajectory moves away from the regime of stable periodic orbits, and then it jumps down to the lower part of Σ, which completes one cycle of action potential.

We note that the existence of spontaneous tonic firing solutions depends on the following two facts: 1) the family of stable periodic orbits of the fast subsystem terminates in a saddle-node on invariant circle (SNIC) and 2) $gtc$-nullsurface lies above the lower part of Σ. If $g_{CaT}$ is decreased, then the $gtc$-nullsurface moves downward and intersects with the lower part of Σ while the family of stable periodic orbits still terminate in a SNIC. The resulting intersection curve is the curve of globally stable fixed points. Hence, the trajectory approaches this curve and cannot enter the regime of stable periodic orbits. As a result, there are no spontaneous tonic firing solutions for small $g_{CaT}$ values.

Two-parameter bifurcation diagram is given in Fig. 2C, which shows RK line (black slant line) with the projection of four spiking solutions for $g_{CaT}$ = 15 (black), 20 (blue), 25 (red), and 40 (magenta). At upper right turning point, trajectory turns counterclockwise. Figure 2C shows that the projection of tonic spiking solution becomes flat and moves rightward as $g_{CaT}$ increases. Based on these two observations, we can provide a heuristic explanation for why the frequency of spiking solution increases as $g_{CaT}$ increases as follows. First, note that $gtc$ increases faster for larger $g_{CaT}$ values because $g_{CaT}$ controls the dynamics of $gtc = g_{CaT} * q$. On the other hand, we may assume that the dynamics of $ghcnc$ is similar in these four examples because $g_{HCN}$ is fixed for all cases ($g_{HCN} = 2$). Thus, the period of spontaneous tonic firing solution may be estimated by the range of $ghcnc$ in the trajectory. Thus, we can expect that the frequency tends to increase and to level off eventually as $g_{CaT}$ increases. Second, recall that, once the trajectory jumps down, then it approaches the lower part of Σ, moves slowly along it, and then eventually crosses the RK line to jump up. In the two-parameter bifurcation diagram, this corresponds to the portion of the trajectory from lower turning point to the upper turning point. In fact, trajectory spends most of time near the lower part of Σ. Now, as $g_{CaT}$ increases, the projected trajectory moves rightward, hence it spends less time near the lower part of Σ due to the proximity to the RK line. In summary, as $g_{CaT}$ increases, trajectory spends less time near the lower part of Σ with slower dynamics, which results in the increase of the spiking frequency.

Now, we investigate why the projection of tonic spiking solution moves rightward as $g_{CaT}$ increases. First, recall that ($gtc$, $ghcnc$, $V$) phase space is divided into two subregions by $gtc$-nullsurface (Fig. 2B). The place where spiking solution is found seems to be determined by the balance between the times that trajectory spends in these two regions while traversing the phase space. Here we note that the position of $gtc$-nullsurface in Fig. 2B depends on $g_{CaT}$ value. In fact, numerical simulation shows that $gtc$-nullsurface moves up as $g_{CaT}$ value increases. Thus, the increase of $g_{CaT}$ value affects the position of spontaneous tonic firing solution near RK.

Since $gtc = g_{CaT} * q$, the position of the spiking solution might be determined by the dynamics of slow variable $q$ in the active phase. Averaging method is frequently used to obtain the reduced autonomous equation that governs the evolution of the slow variable in the active phase. Formally, if $\frac{dx}{dt} = f(x, t)$ is the system for the evolution of slow variable $x$ and $f(x, t)$ is of period $T$, then the associated autonomous averaged system in the active phase is given by

$$\bar{x} = \frac{1}{T} \int_0^T f(x, t)\, dt := \overline{f(\bar{x})}$$

, which describes the dynamics of slow variable in the active phase. Now, $\frac{dq}{dt} = \frac{q_\infty(V)-q}{\tau_q(V)}$, hence the position of the spiking solution might be determined by the averaged value of $q_\infty(V)$. In this case, since the time constant $\tau_q(V)$ also depends on the voltage, we used a weighted averaging method for $q_\infty(V)$ in the regime of stable periodic orbits. More precisely, we computed the following averaged value of $q_\infty(V)$ in a periodic regime as

$$\overline{q_\infty} = \int_0^T \frac{q_\infty(V)}{\tau_q(V)} dt \Big/ \int_0^T \frac{1}{\tau_q(V)} dt$$

where $T$ is the period of a stable periodic orbit. That is, for a fixed $ghcnc$ value, we draw a bifurcation diagram and find $gtc$ value for SNIC, say $gtc_{SNIC}$. From $gtc_{SNIC}$, we consider 30 points of $gtc$ with step size 0.001. Here, each fixed $gtc$ value corresponds to a periodic orbit. Now we computed averaged $\overline{q_\infty}$ over this periodic orbit and Fig. 2D shows the averaged $gtc$ ($\overline{gtc} = g_{CaT} * \overline{q_\infty}$), when $ghcnc = 0.01$. The results for other $ghcnc$ values show qualitatively similar patterns. We checked three $g_{CaT}$ values, 20 (blue), 30 (red), and 40 (magenta). Diagonal solid line is the line of identity. Since $\overline{gtc}$ denotes an equilibrium value of $gtc$ over a spiking solution, if $\overline{gtc}$ line lies above (below, resp.) the line of identity, then $gtc$ is forced to increase (decrease, resp.). Thus, the intersection between the line of identity and $\overline{gtc}$ curve denotes the $gtc$ value where spontaneous tonic firing solution tends to reside. As shown in Fig. 2D, the intersection point moves rightward as $g_{CaT}$ increases and this result explains why the projected trajectory in Fig. 2C moves rightward as $g_{CaT}$ increases. We also note that when $g_{CaT} = 40$ in Fig. 2D, the intersection point is around 0.125 and Fig. 2C shows that the value lies on the right side of RK line. This suggests that the spontaneous tonic firing solution in this case may lie inside the regime of stable periodic orbits and the trajectory spends almost no time on the lower surface of Σ which results in a higher frequency.

## 3.2 Effect of the CaL current on spontaneous tonic firing. Bifurcation analysis.

In this section, we present the effect of the CaL current on spontaneous tonic firing by varying $g_{CaL}$ values (default value is 5) with fixed $g_{CaT}$ and $g_{HCN}$ values. We tested six different $g_{CaL}$ values (5, 10, 15, 20, 30, and 40). Over these $g_{CaL}$ values, we found that the model yields low frequency spiking solutions when $g_{CaL} \leq 15$ and high frequency spiking solutions when $g_{CaL} \geq 20$ (Fig. 1D). As seen in Fig. 1D, the frequency of the spiking solution was not affected by the HCN current even if the timescale of the dynamics of the HCN current is substantially increased. That is, the timescale of kinetics of the HCN current may not play a significant role when the CaL current is not negligible anymore. For this reason, we utilized the following slow variables for bifurcation analysis: $gtc = g_{CaT} * q$, $gcalc = g_{CaL} * d_1 * d_2$, and $[Ca]$, where $d_1$ and $d_2$ are slowly activating/de-activating gating variables in the CaL current, and $[Ca]$ is the calcium concentration. We would like to note that the numerical simulation shows that average calcium level $[Ca]$ is around 0.05 in low frequency spiking solutions and is around 0.18 in high frequency spiking solutions.

Figure 3A and 3B show bifurcation diagrams of the fast subsystem in $(gtc, V)$-space with the bifurcation parameter $gtc$ for fixed $gcalc$ and $[Ca]$ values. Here, $[Ca]$ value is 0.05 in Fig. 3A and 0.18 in Fig. 3B. In each figure, $gcalc$ values are 2 (black), 6 (blue), and 10 (red). Green curves are $gtc$-nullcline when $g_{CaT} = 20$. There are several things to note on these bifurcation diagrams. First, we see that the lower and middle parts of the S-shaped curve of fixed points ($S$) remain almost the same over various $gcalc$ values. Especially the dependence of $gtc$ value of the right knee (RK) of $S$ ($gtc_{RK}$) on $gcalc$ values is negligible. This is clearly shown in Fig. 3C, where the vertical lines denote the RK lines for $[Ca]$ = 0.05 (black) and $[Ca]$ = 0.18 (blue). Second, there are two different termination mechanisms of stable periodic orbits depending on the $gcalc$ values. As shown in the previous section, the family of stable periodic orbits emanates from $S$ at the AH-point. For smaller $gcalc$ values such as 2, stable periodic orbits terminate in a saddle-node on an invariant circle (SNIC) bifurcation. On the other hand, for larger $gcalc$ values such as 6 and 10, the stable periodic orbits turn around at saddle-node bifurcation of periodic orbits (SNPO) to become unstable periodic orbits. The third thing to note is the relative position of $gtc$-nullcline with respect to $S$, especially with respect to the RK of $S$. As seen in the figure, $gtc$-nullcline lies above the RK for smaller $[Ca]$ values (Fig. 3A). But $gtc$-nullcline intersects at the middle and lower parts of $S$ for larger $[Ca]$ values (Fig. 3B).

The former case is similar to the one in the previous section, which implies that there will be a spontaneous tonic spiking solution near the RK for smaller $[Ca]$ values. For larger $[Ca]$ values, on the other hand, a different mechanism comes in. Note that the intersection between $gtc$-nullcline and the S-shaped curve on the lower branch of stable fixed points is a globally stable fixed point. Hence, if a trajectory jumps down to the lower branch of $S$, then it will approach this globally stable fixed point along the lower branch and remain there. Thus, there is no spiking solution near the RK. The other possible way to obtain a spiking solution is inside the regime of stable periodic orbits if the averaged $gtc$ value ($\overline{gtc}$) is somewhere between the SNPO and the RK. Since the trajectory does not lie on the lower stable fixed points and does not stay in the periodic orbit, the frequency is higher than ones near RK (Fig. 3A).

Figure 3C shows a two-parameter bifurcation diagram with the projection of full model spiking solutions. The two vertical lines in the figure are the lines of the RK for $[Ca]$ = 0.05 (black) and 0.18 (blue). As shown in Fig. 3A-B, these lines are almost independent of $gcalc$ values. Dotted lines that emanate from the RK lines are SNPO lines. To obtain these SNPO lines, we checked the $gtc$ values of SNPO for $gcalc$ = 1, 2, 3, …,10, and then connected these points. Thus, the SNPO curves are not smooth enough at some places. Black horizontal lines on the black SNPO line denote the projection of full model spiking solutions for $g_{CaL}$ = 5, 10, and 15 (from bottom to top). The short horizontal lines at the upper left part denote the projection of full model spiking solutions for $g_{CaL}$ = 20 (black), 30 (blue), and 40 (red). We can see that spiking solution resides near the RK for small $g_{CaL}$ values and occurs away from the RK line for large $g_{CaL}$ values. This suggests that for small $g_{CaL}$ values, the trajectory jumps down to the lower branches of the stable fixed points and approaches to RK line while for large $g_{CaL}$, the trajectory does not jump down to the lower branches of the stable fixed points.

Figure 3D shows a bifurcation surface in ($gtc$, $gcalc$, $V$)-space when $[Ca]$ = 0.18 with the projection of tonic spiking solutions for $g_{CaL}$ = 20 (black), 30 (blue), and 40 (red). Minimum and maximum values of $V$ along the family of periodic orbits form black surfaces in the figure, which are folded at the SNPO lines. Inner parts correspond to stable periodic orbits and outer parts to unstable periodic orbits. The S-shaped surface of fixed points (Σ) is shown in cyan. Figure 3D shows that spiking solutions exist inside the regime of stable periodic orbits. Since slow variables ($gtc$, $gcalc$) are not slow enough, spiking solutions are not confined inside the region between two surfaces of stable periodic orbits. As $g_{CaL}$ decreases, spiking solution approaches the SNPO line.

Figure 3E shows the averaged $gtc$ value ($\overline{gtc}$) for various $[Ca]$ and $gcalc$ values. Recall that the intersection between the averaged $\overline{gtc}$ and the line of identity is where spiking solution tends to reside. The upper two curves are for $gcalc$ = 2 and 4 when $[Ca]$ = 0.05. These cases are similar to those in previous sections so that a spiking solution tends to reside near the RK. The lower two horizontal curves are for $gcalc$ = 8 and 10 when $[Ca]$ = 0.18. This result shows that spiking solution tends to reside inside a periodic regime. Since $[Ca]$ increases while voltage goes up, the overall level of $[Ca]$ depends on the frequency of spiking solution. When spiking solution is near the RK, the spiking frequency is low, hence, the overall level of $[Ca]$ is low, too. Therefore, when $[Ca]$ = 0.05 (Fig. 3A), $\overline{gtc}$ values at upper right corner have values close to what is expected (Fig. 3E). On the other hand, when $[Ca]$ =0.18 (Fig. 3B), $\overline{gtc}$ have values at lower left corner (Fig. 3E).

In summary, when $g_{CaL}$ is relatively small, the model yields a spontaneous tonic spiking solution, which is similar to those shown in previous section and is facilitated by the CaT current. But, when $g_{CaL}$ is sufficiently large, then the CaL current pushes the trajectory into the regime of stable periodic orbits, hence there is an abrupt jump in the frequency during this transition and we obtain higher frequency spiking solutions. In terms of a bifurcation diagram, the availability of the CaL current pushes a bifurcation diagram upward, hence the RK of S-shaped curve $S$ lies above the $gtc$-nullcline. This change does not allow for a spontaneous tonic spiking solution near the RK but creates a solution inside the regime of stable periodic orbits.

## 4. Hyperpolarization-induced bursting rhythms

Experimental and computational studies point to the importance of CaT and CaL currents for the generation of hyperpolarization-induced bursting rhythms in STN. Beurrier et al. [Beurrier et al., 1999] showed that some STN neurons can switch from the spontaneous tonic firing to slow bursting rhythms or mixed burst-firing patterns under a sustained hyperpolarizing current application. They argued that CaT and CaL currents underlie the generation of the slow rhythmic bursting. Gillies and Willshaw [Gillies and Willshaw, 2006] showed that their multi-compartment model generates a slow rhythmic bursting in the presence of a uniform reduction in the Ca-dependent SK conductance (simulating the application of apamin) and constant hyperpolarizing current injection. They argued that the interaction of CaT and CaL currents determines the presence and nature of the rhythmic bursting. They also argued that a sufficiently strong CaT current was necessary for the generation of individual bursts. Our model is also able to generate bursting rhythms under a sustained hyperpolarization. Similar to the model in [Gillies and Willshaw, 2006], it was necessary to reduce the AHP current conductance and increase the CaT current conductance to generate bursting rhythms. In this section, we study the effect of CaT and CaL currents as well as the HCN current on hyperpolarization-induced bursting rhythms via bifurcation analysis.

### 4.1 Effect of CaT and HCN currents on hyperpolarization-induced bursting rhythms. Bifurcation analysis.

First, we will explore how CaT and HCN currents in an STN neuron under hyperpolarization generate bursting rhythms without the CaL current. The default parameter values in this case are $g_{CaT} = 25$, $g_{AHP} = 0.2$, $g_{CaL} = 0$, $g_{HCN} = 2$, and $I_{app} = -16$. Note that $I_{app}$ is a large (in terms of the magnitude) negative number to generate hyperpolarization-induced bursting rhythms.

Figure 4A-B show bursting rhythms for different values of $g_{CaT}$ with fixed $g_{HCN}$ (Fig. 4A) and for different values of $g_{HCN}$ with fixed $g_{CaT}$ (Fig. 4B). As $g_{CaT}$ increases, the period decreases while burst duration and number of spikes within a burst increase (Fig. 4C). Thus, the inter-burst interval decreases too. On the other hand, as $g_{HCN}$ increases, we observed that period, burst duration, and number of spikes within a burst decrease at the same time (Fig. 4C). Please, note that the checkerboard pattern shown in the middle figure (burst duration) is due to the typical spike-adding procedure which is frequently found in bursting regime with a small number of spikes within a burst.

To explore the underlying mechanisms of these results, we performed a bifurcation analysis using $gtc$ and $ghcnc$ (as defined in the previous section: $gtc = g_{CaT} * q$ and $ghcnc = g_{HCN} * f$). Figure 4D shows bifurcation diagrams of the fast subsystem projected onto ($gtc$, $V$)-space with a bifurcation parameter $gtc$ for $ghcnc$ = 0.1 (black) and 0.2 (red). Green curve is the $gtc$-nullcline for $g_{CaT}$ = 25. While bifurcation structures are qualitatively similar to those shown in previous section, there are some important characteristic differences compared to those in the previous section. First, the family of stable periodic orbits lies above $gtc$-nullcline. This holds true for all reasonable $ghcnc$ values. Hence, when the projection of full model solution is in a bursting mode, or in other words, when the projected trajectory is inside the regime of stable periodic orbits, the trajectory moves leftward and eventually jumps down to the lower branch of the S-shaped curve of the stable fixed points ($S$) at the homoclinic (HC) point. Second, bifurcation diagrams show that middle and lower parts of $S$ move leftward (to the lower values of $gtc$) as $ghcnc$ increases. When $ghcnc$ is small, there is an intersection point between $gtc$-nullcline and the lower branch of $S$, which is globally stable. As $ghcnc$ increases, this intersection point approaches RK of $S$. Hence, when $ghcnc$ is small, if the trajectory is in the small neighborhood of the lower branch of $S$, then it moves along the lower branch of $S$ to approach this intersection point. On the other hand, while approaching, $ghcnc$ value increases, which results in the loss of this globally stable fixed point. Then, the trajectory is able to jump up into the stable periodic orbit regime and spiking within a burst begins.

Figure 4E shows bifurcation surfaces in ($gtc$, $ghcnc$, $V$)-space with bifurcation parameters $gtc$ and $ghcnc$. Projection of bursting solution for $g_{CaT}$ = 25 and $g_{HCN}$ = 2 is also shown (red). The $gtc$-nullsurface is omitted and the S-shaped surface of fixed points ($\Sigma$) is shown in cyan for the clarity of the figure. The projection of bursting solution is not confined inside the regime of stable periodic orbits since slow variables ($gtc$, $ghcnc$) are not sufficiently slow. If we make $gtc$ and $ghcnc$ slower, then we can obtain bursting solutions with a longer burst duration and a larger number of spikes, which are confined inside the regime of stable periodic orbits. However, current bifurcation diagrams are sufficient to analyze hyperpolarization-induced bursting mechanisms. Figure 4E shows that the projected trajectory moves along the lower part of $\Sigma$; both $gtc$ and $ghcnc$ values increase. Once the trajectory crosses the RK line, it jumps up into the regime of stable periodic solutions. Since the family of stable periodic orbits lies above the $gtc$-nullsurface, $gtc$ and $ghcnc$ keep decreasing while the trajectory is inside the stable periodic orbit regime. The trajectory eventually jumps down to the lower part of $\Sigma$, which completes one cycle of a bursting solution.

Figure 4F shows a two-parameter bifurcation diagram with projection of full model bursting solutions for different values of $g_{CaT}$ with fixed $ghcnc$ (all values are the same as in the Fig. 4A with time-series). Recall that both $gtc$ and $ghcnc$ increase over the silent phase of bursting solution and this corresponds to the almost straight, increasing part of the projected trajectory. Once it passes the RK line, the trajectory turns around counterclockwise to start spiking. Wiggles of the projected trajectory near its lower left part correspond to the active phase of a burst. If the trajectory crosses the HC line, it jumps down to the lower branch of stable fixed points and the active phase of a burst is terminated. Note that the active phases of burst in all three trajectories terminate at similar places (cf. Fig. 4F). As $g_{CaT}$ increases, the projected trajectory is stretched horizontally and it moves down rightward, hence the range of $gtc$ increases, whereas the range of $ghcnc$ decreases. We may assume that $ghcnc$ evolves on similar timescales in all three bursting solutions when $g_{HCN}$ is fixed. Thus, the period and the interburst interval of the bursting solution can be estimated by the range of $ghcnc$. More specifically, they can be estimated by the maximum value of $ghcnc$ where the trajectory turns around to begin the active phase of a burst in the figure. Since this maximum value of $ghcnc$ decreases as $g_{CaT}$ increases, we expect that the period and the interburst interval also decrease as $g_{CaT}$ increases.

A similar argument can be applied to the bifurcation diagram in Fig. 4G, which shows the projection of a full model bursting solutions for different values of $g_{HCN}$ and fixed $g_{CaT}$ (all values are the same as in Fig. 4B with time-series). In this case, as $g_{HCN}$ increases, the projected trajectory is stretched vertically and moves up leftward. Thus, the range of $gtc$ decreases whereas the range of $ghcnc$ increases. When $g_{CaT}$ is fixed, the interburst interval and the period can be estimated by the range of $gtc$ since we may assume that $gtc$ evolves on similar timescales in all three bursting solutions. Therefore, the interburst interval and the period decrease as $ghcnc$ increases because the range of $gtc$ decreases as $ghcnc$ increases.

The burst duration and the number of spikes within a burst are determined by the proximity of the trajectory to the AH line. Note that proximity to the AH line means there is a higher chance to fire because all trajectories terminate at similar places for different values of $g_{CaT}$ and $g_{HCN}$ (Fig. 4D-G). As a result, the burst duration, and the number of spikes within a burst increase in both cases when trajectory moves rightward or more precisely, as $g_{CaT}$ increases with fixed $ghcnc$ and as $ghcnc$ decreases with fixed $g_{CaT}$.

Figure 4F,G also tells us about the role of the HCN current. Without the HCN current ($ghcnc$ = 0), the CaT current driven bursting solution is not possible. Recall that the active phase of a burst is initiated when the trajectory crosses the RK line. Without the HCN current, the projection of solution trajectory approaches the globally stable fixed point on the lower branch of the fixed points (Fig. 4D), hence the trajectory cannot cross the RK line to initiate a burst. In two-parameter bifurcation diagrams, the lack of the HCN current means that the projected trajectory lies on the horizontal axis (Fig. 4F-G). Similarly, we can argue that small HCN current means a very long interburst interval. Also, note that the number of spikes within a burst is determined by the proximity to the AH line. Consequently, due to the shape of the AH line, the number of spikes within a burst will be capped. Large HCN currents would result in high frequency spiking since the interburst interval and the number of spikes within a burst decrease at the same time. This result concurs with the experimental observations showing that HCN channels promote single-spike activity rather than bursting rhythms [Atherton et al., 2010].

## 4.2 Effect of the CaL current on hyperpolarization-induced bursting rhythms. Bifurcation analysis.

In this section, we studied the effect of the CaL current on the hyperpolarization-induced bursting rhythms. Default parameter values are $g_{CaT} = 25$, $g_{AHP} = 0.2$, $g_{CaL} = 15$, $I_{app} = -22$, $g_{HCN} = 2$. Figure 5A shows voltage profiles of bursting solutions for different values of $g_{CaL}$. Here, we found that burst duration increases substantially as $g_{CaL}$ increases, while interburst interval increases slightly. These results are summarized in Fig. 5B, which shows period (solid line), interburst interval (dashed line), and burst duration (dotted line) as a function of $g_{CaL}$. While both burst duration and interburst interval increase with $g_{CaL}$, the increase of burst duration is more significant and so is the period. As compared to CaT current bursting solutions considered in the previous section, the burst duration and the number of spikes within a burst are substantially increased when the CaL current is turned on. Figure 5C,D shows period and burst duration over two-parameter space for $g_{CaL}$ = 5, 15, and 25. As shown in Fig. 5C, the region for large periods increases as $g_{CaL}$ increases.

To explore the underlying mechanisms that result in these differences, we performed bifurcation analysis using $gcalc = g_{CaL} * d_1 * d_2$, $gtc = g_{CaT} * q$, and $[Ca]$. We did not choose $ghcnc$ as a slow variable although $ghcnc$ is slow over silent phase of bursting solution. In fact, if we increase the timescale of the HCN current, we still obtain qualitatively similar results. This may be because the HCN current is mostly involved in the transition from the silent phase to the active phase of bursting solution while the CaL current affects the active phase of the bursting solution more significantly. Figure 5E shows two exemplary bifurcation diagrams with bifurcation parameter $gtc$ for $[Ca]$ = 0.4 and $gcalc$ = 5 (dotted), and $[Ca]$ = 0.4 and $gcalc$ = 10 (solid). The green curve is a $gtc$-nullcline for $g_{CaT}$ = 25. Similar to the bifurcation diagrams shown in Fig. 3, we observed that 1) the lower and middle parts of S-shaped curve of fixed points ($S$) remain almost the same over various $gcalc$ values, 2) the AH-point moves leftward as $gcalc$ increases, and 3) the stable periodic orbits turn around at saddle-node bifurcation of periodic orbits (SNPO) to become unstable periodic orbits. On the other hand, there are two important differences to note between these diagrams and the ones in Fig. 3. First difference is the relative position of $gtc$-nullcline with respect to the stable periodic orbits. Bifurcation diagrams in this section show that the branches of stable periodic orbits lie above $gtc$-nullcline for small $gcalc$ values. In this case, bursting solution keeps moving leftward over the active phase of a bursting solution. For large $gcalc$ values, the stable periodic orbits intersect with $gtc$-nullcline and the averaged $gtc$ ($\overline{gtc}$) will have a value close to $gtc$ value at SNPO (cf. Fig. 3). The second difference is the proximity of SNPO to $gtc$-nullcline for large $gcalc$ values. This proximity forces the bursting solution to slow down while approaching SNPO over the active phase of bursting solution. These two facts imply that bursting solution slows down while approaching $\overline{gtc}$ value near SNPO and contribute to longer active phase of bursting for large $gcalc$ values.

Figure 5F shows SNPO surface (left), the AH surface (middle slant), and the RK surface (right) in ($gtc$, $gcalc$, $[Ca]$)–space. The same figure also shows projections of the three bursting solutions for different values of $g_{CaL}$. Note that large $g_{CaL}$ values mean there is an elevated range of $gcalc$ values. Over the silent phase of a burst, $gtc$ and $gcalc$ increase while $[Ca]$ decreases. In this model, $[Ca]$ decreases sufficiently fast so that the RK line when $[Ca] = 0$ roughly determines when the cell crosses the RK line and enters the regime of stable periodic orbits. Similar to the bifurcation diagrams in Section 3.2, we see that the RK surface is almost vertical with fixed $[Ca]$ value, in other words, almost independent of $gcalc$ values. This fact implies that the bursting solution is facilitated by the CaT current. Once the cell enters the regime of stable periodic orbits, $gtc$ and $gcalc$ begin to decrease while $[Ca]$ begins to increase initially and then decrease over an active phase of burst. Spiking within a burst is terminated once trajectory crosses the surface of SNPO. Since burst duration is determined by the distance between the RK surface and the SNPO surface, burst duration increases as $g_{CaL}$ increases. Especially because, as evidenced by Fig. 5E, bursting solution spends longer time near the SNPO before jumping down for larger $gcalc$ values, thus there is even longer bursting duration and large number of spikes within a burst for large $gcalc$ values.

## 5. Post-inhibitory rebound burst

In this section we study calcium-dependent post-inhibitory rebound (PIR) bursts of spikes when the model neuron is released from application of inhibitory current. This section will cover 1) general mechanism underlying PIR, 2) the effect of magnitude and duration of inhibitory current application, and 3) the effect of CaT and CaL currents on PIR. The default parameter values for this section are $g_{CaT} = 20$, $g_{AHP} = 1$, $g_{CaL} = 5$, $g_{HCN} = 2$. For the simulation of inhibitory input, we used the parameter $I_{app}$ in the differential equation of membrane potential $V$ (Eq. 1). Default magnitude and duration of applied inhibitory current are -20 (that is $I_{app} = -20$) and 500ms.

### 5.1 Mechanisms underlying PIRs. Bifurcation analysis.

Figure 6A shows an example of PIR. Before a cell is given an inhibitory input, the cell exhibits a spontaneous tonic firing with a frequency of around 10 Hz. At *t* = 500 msec, $I_{app}$ was changed from 0 to -20 for 500ms and then the cell was hyperpolarized over this period. Once the inhibitory input was removed at *t* = 1000 msec, the cell exhibited a burst where the frequency decreased over time while the magnitude increased. Although there is no clear way to define the duration of PIR, we can loosely define it as the time from the removal of inhibition to the moment when the trajectory returns to its original tonic firing solution (Fig. 6E-F). In terms of inter-spike intervals, this means that inter-spike interval returns to its original value in spontaneous tonic firing solution. In Fig. 6A, the trajectory returned to its original spontaneous tonic firing around t = 1600ms, thus the duration of PIR is around 600ms in this example.

In previous sections we observed that bursting solutions and spiking solutions are facilitated by the CaT current and modulated by the HCN-current when the CaL current is less significant. Therefore, we chose $gtc$ and $ghcnc$ as bifurcation parameters and studied the underlying mechanism of PIRs. Figure 6B-D show bifurcation surfaces in ($gtc$, $ghcnc$, $V$) space with the projection of full model PIR solution shown in Fig. 6A. As before, green surface denotes $gtc$-nullsurface for $g_{CaT} = 20$. Figure 6B shows the spontaneous tonic spiking solution until *t* = 500ms, which is similar to the one in Fig. 2B. Figure 6C illustrates what happens during the constant inhibitory input. Under inhibition, the surface of fixed points (Σ) moves rightward and as a result, the RK line of Σ is also shifted rightward. Then, the trajectory jumps down to the lower part of Σ (lower white surface with black grid lines in the figure) and moves along the surface. Recall that $gtc$ and $ghcnc$ increase at the same time along the surface. Let Γ denote the intersection curve of lower part of Σ and $gtc$-nullsurface. This curve Γ is the set of globally stable fixed points and shown in the figure as a red line on the lower part of Σ. Since Γ lies on the lower surface of Σ, the trajectory (blue) moves along the lower surface of Σ to approach Γ and stays there until the inhibition is removed. In other words, this intersection curve Γ holds the trajectory until the removal of the inhibition and delimits the maximum level of $gtc$ and $ghcnc$ during the inhibition. Hence, PIR duration is also delimited by this curve Γ. Once the inhibition is removed, the RK line goes back to its original place, the trajectory jumps up into the regime of stable periodic orbits, and a burst begins (Fig. 6D). While spiking within a burst, both $gtc$ and $ghcnc$ decrease but $ghcnc$ decreases faster at the beginning of the burst. When the cell is released from inhibition, trajectory is close to the AH point so that the burst frequency is high. As the trajectory traverses the regime of stable periodic orbits toward the end of the regime (SNIC) on the left, the frequency of PIR decreases. After crossing the RK line, which is identical to the SNIC line (cf. Fig.2), the trajectory approaches tonic spiking solution (Fig. 6B).

Figure 6E shows a two-parameter bifurcation diagram with the projection of the PIR solution (blue). The RK line is also shown as a black slant line. Thick blue curves near RK line correspond to spontaneous tonic spiking before the inhibition and after PIR. Once the inhibition is turned on, $gtc$ and $ghcnc$ begin to increase (blue diagonal line). Figure 6F shows what happens during and after the inhibition. Figure 6E corresponds to the lower left corner of Fig. 6F. Once the inhibition is turned on, RK line is shifted right upward (black dotted line on the upper right corner; the red line denotes Γ,cf. Fig. 6C). Now the trajectory approaches Γ. Once the inhibition is removed, the RK line goes back to its original place (black solid line on the left side) and the trajectory jumps up into the regime of stable periodic orbits and a burst begins (cf. Fig. 6D). While spiking within a burst, the trajectory approaches the spontaneous tonic firing solution (cf. Fig. 6B,E).

**5.2 Effect of magnitude and duration of inhibitory input on PIRs. Bifurcation analysis.**

In this section, we studied the effect of magnitude and duration of inhibitory input on PIRs using a two-parameter bifurcation diagram. Figure 7A shows voltage profiles for progressively larger values of negative $I_{app}$ ($I_{app}$ = -5, -10, -20, and -30 from top to bottom). The duration of inhibitory input is 500 ms for all four cases. As the magnitude of inhibitory input increases (input is more hyperpolarizing), the duration of PIR also increases. For more hyperpolarizing inputs such as $I_{app}$ = -30, PIR shows an initially high frequency response and then the frequency of spiking slowly goes down to its base value. Figure 7B shows a two-parameter diagram with the projection of solutions for the same values of $I_{app}$ as the time-series in Fig. 7A. The black curve at lower left corner is the RK line for $I_{app} = 0$. The remaining thin curves are intersection curves (Γs) described in previous section. These intersection curves (Γs) follow the same color code as the projected trajectories. As the magnitude of inhibitory input increases, Γ shifts to the upper right corner. When $I_{app} = -5$ or $-10$, input duration 500 ms was long enough so that trajectory reached Γ and stayed there. When $I_{app} = -20$ or -30, on the other hand, inhibitory input is effectively removed while the trajectory is still approaching Γ. Although there are two different scenarios, depending on magnitude and duration of inhibitory input, we see that, as the magnitude of inhibitory input increases, trajectory traverses farther away from the RK line (when $I_{app} = 0$) and the duration of PIR increases as a result. This fact also implies that, for large magnitude inhibitory input, a trajectory is closer to the AH point when it is released from inhibition. In other words, a large magnitude inhibitory input means a proximity to the AH point. Thus, we observe high frequency spiking at the beginning of PIR for large magnitude inhibitory input.

Figure 7C shows voltage profiles with four different input durations. Here $I_{app}$ is fixed at -20. As the input duration increases, the duration of PIR increases. We also observe that two PIRs for input duration 500ms and 750ms look similar. Figure 7D shows a two-parameter diagram with the projection of solutions when input durations are the same as for the time-series in Fig. 7C. Termination (removal) of inhibitory input is denoted by dot in each trajectory. In all cases, trajectories approach the intersection curve Γ (thin magenta) over inhibition. When duration is 750ms, trajectory crosses Γ but it is trapped by that curve while $ghcnc$ increases. Hence trajectories move along Γ until inhibition is terminated. Figure 7D illustrates that PIR durations for input duration 500ms and 750 ms are similar because $gtc$ values at release are similar.

**5.3 Effect of CaT and CaL currents on PIRs. Bifurcation analysis.**

Figure 8A shows the effect of $g_{CaT}$ on PIRs using three $g_{CaT}$ values, 15, 20, and 30 (from top to bottom) with fixed $g_{CaL} = 5$, input duration 500 ms, and $I_{app} = -20$. As shown in Fig. 1, the frequency of spontaneous tonic firing increases as $g_{CaT}$ increases. Burst duration in PIR shows a slight but not significant increase as $g_{CaT}$ increases. In the two-parameter diagram (Fig. 8B), the increase of $g_{CaT}$ results in the horizontal stretch of the projected trajectories. The intersection curves (Γs) are also shifted to the right. This figure shows that maximum levels of $ghcnc$ values are similar in all cases. Since PIR duration can be estimated by the maximum level of $ghcnc$ when $g_{HCN}$ is fixed, this explains why there is only a minor change in PIRs as $g_{CaT}$ increases. When $g_{CaT} = 30$, the trajectory is close to the AH point when it is released from inhibition. Due to this proximity to the AH point, PIR shows high frequency and small-magnitude spiking at the beginning of PIR (cf. Fig. 6).

Now, we consider the effect of the CaL current on PIRs. Figure 8C shows PIRs for $g_{CaL}$ = 5, 10, and 15 (from top to bottom) with fixed $g_{CaT} = 20$, input duration 500 ms, and $I_{app} = -20$. As $g_{CaL}$ increases, we observe that the duration of PIR increases substantially. Note that the dynamics of the trajectory until jumping up into the regime of stable periodic orbits is facilitated by the CaT current and the HCN-current as discussed in Sections 3 and 4. In fact, when a cell is either in spontaneous tonic firing or hyperpolarized by inhibitory input, $[Ca]$ is almost constant at a very low level. Consequently, the activity patterns of these two states (either spontaneous tonic firing state before inhibitory input or silent state under hyperpolarizing input) can be explained by two slow variables $gtc$ and $ghcnc$ as before. This is confirmed by Fig. 8D, which shows the projection of three solutions (for different $g_{CaL}$ values) and the corresponding intersection curves in $(gtc, ghcnc)$-plane when cell is hyperpolarized. Both the projected trajectories and intersection curves for different values of $g_{CaL}$ ($g_{CaL}$= 5, 10, 15) are not distinguishable. In other words, either the spontaneous tonic firing state before inhibitory input or silent state under hyperpolarizing input do not depend on $g_{CaL}$ values when $g_{CaL}$ is small.

However, when a cell is released from inhibition and jumps into the regime of stable periodic orbits, the CaL current begins to play an important role in PIR. We can explain the dynamics of trajectory over PIR using $[Ca]$, $gcalc$, and $gtc$. Figure 8E shows the projection of trajectories in $(gtc, gcalc, [Ca])$ -space for $g_{CaL}$ = 5 (blue), 10 (red), and 15 (magenta). We omitted the RK surface and the AH surface, and plotted the SNPO/SNIC surface only in the middle for clarity (cf. Fig. 5D). Figure 8F shows the same trajectories with a SNPO/SNIC curve when $[Ca]$ = 0.05. Once a cell is released from the inhibitory input, $[Ca]$ initially increases ,and then overall level of $[Ca]$ decreases over spiking within a burst. The overall decrease of $[Ca]$ level is due to the decrease of spiking frequency over the time course of the active phase of burst. The values of $gtc$ and $gcalc$ also decrease over spiking within a burst. For small $g_{CaL}$ values, for example, $g_{CaL} = 5$ (red) and 10 (blue), $gcalc$ values are also relatively small (around 2.5 when $g_{CaL} = 5$ and around 5 when $g_{CaL} = 10$). In these two cases, active phase of burst is terminated when trajectory crosses the SNPO surface (Fig. 8F). Recall that in Fig. 5, we showed that the branches of stable periodic orbits lie above the $gtc$-nullcline, hence the bursting solution moves leftward and jumps down to the lower branch of the S-shaped curve of fixed points $S$. After jumping down, in this case, trajectory approaches spontaneous tonic firing solution (Fig. 8F). On the other hand, if $g_{CaL}$ is sufficiently large, the bursting solution is not terminated by the SNPO because the SNPO has a negative $gtc$ value over large $gcalc$ values (Fig. 8F). As shown in Fig. 3, since the lower end part of the branch of stable periodic orbits including the SNPO lies below $gtc$-nullcline, averaged $gtc$ ($\overline{gtc}$) is between 0 and the RK. Thus, bursting solution moves leftward to approach $\overline{gtc}$. And while doing so, the $[Ca]$ level keeps decreasing and $\overline{gtc}$ increases slowly (Fig. 3E). Figure 8F shows that the trajectory for $g_{CaL} = 15$ (magenta) turns around while spiking. Now the $[Ca]$ level becomes sufficiently low and the $gcalc$ value is sufficiently small. Then the PIR dynamics undergoes a transition to the dynamics of spontaneous tonic firing, which is driven by the CaT current and the HCN-current.

# 6   Discussion

In this study, we presented a conductance-based single-compartment model of an STN neuron, which plays an important role in the pathophysiology of the basal ganglia in Parkinson's disease. STN neurons exhibit characteristic activity patterns such as: a slow rhythmic firing [Beurrier et al., 1999; Bevan and Wilson, 1999], a calcium dependent post-inhibitory rebound (PIR) bursts [Bevan et al., 2002a], and slow rhythmic bursting under sustained hyperpolarization [Beurrier et al., 1999]. Recent experiments showed that interaction between the T-type calcium (CaT) current and the L-type calcium (CaL) current plays an important role in the generation of STN activity patterns [Beurrier et al., 1999; Bevan and Wilson, 1999; Bevan et al., 2002a].

The first single-compartment model of an STN neuron was developed by Terman et al. [Terman et al., 2002] and this model was able to generate PIR bursts with the CaT current. Gillies and Willshaw [Gillies and Willshaw, 2006] developed a multi-compartment model, which contained the CaT current, the CaL current, and the HCN current and was able to generate characteristic activity patterns as stated above. However, the interaction of compartments in the model appears to be essential for its dynamical regimes. On the other hand, Hahn and McIntyre [Hahn and McIntyre, 2010] developed a single-compartment model that contained the CaT current and the CaL current but this model does not exhibit PIR burst nor slow rhythmic bursting under sustained hyperpolarization.

The STN model in this study is, to the best of our knowledge, the first single-compartment STN model that is able to generate characteristic activity patterns of STN neurons, especially activity patterns under hyperpolarization (hyperpolarization-induced bursts and PIR bursts). To investigate the roles and effects of these currents in rhythm generation we performed a bifurcation analysis using slow variables. We found that 1) the HCN current promotes single-spike activity patterns rather than bursting rhythms while nonetheless being an essential component for the bursting rhythms, 2) the CaT current enables STN cells to display characteristic firing patterns under hyperpolarization (hyperpolarization-induced bursts and PIR bursts), and 3) the CaL current enriches and reinforces these bursting rhythms under hyperpolarization and PIR.

**6.1 Roles of HCN, CaT, and CaL currents**

Experimental results showed that the HCN current promotes single-spike activity patterns rather than bursting rhythms [Atherton et al., 2010]. The bifurcation analysis of our model showed that the increase of maximal conductance $g_{HCN}$ (making the HCN current stronger) tends to yield a higher chance for spiking solution (Fig. 4G). This fact resulted from the proximity of a trajectory (projection of full model solution onto bifurcation diagram) to the RK line in the bifurcation diagram (cf. Fig. 2A,C, Fig. 4F,G). The RK line is the set of fold bifurcation points where the lower branch of stable fixed points turns around to become the middle branch of the unstable fixed points in the bifurcation diagram. Easier access to the RK line yields a higher chance for obtaining spiking solution in general. In a spontaneous tonic spiking solution, for example, the proximity of the trajectory to the RK line results in shorter inter-spike intervals and a higher frequency of spiking solution (Fig. 2C). Similarly, in a hyperpolarization-induced bursting solution this results in a higher frequency of bursting solution with fewer spikes within a burst (Fig. 4B,G). In conclusion, the larger availability of the HCN current renders means there is an easier access to the RK line, which facilitates a tonic spiking solution.

Our model showed that the CaT current is necessary for activity patterns under hyperpolarization (hyperpolarization-induced burst or post-inhibitory rebound (PIR) burst). This fact, in turn, indicates that the CaT current enables STN cells to generate various firing patterns under hyperpolarizing stimuli within the basal ganglia. The blocking or disrupting the CaT current may mute the emergence of rebound responses and hyperpolarization-induced rhythmic bursting solution. Our model also shows that the addition of the CaL current makes the response of an STN cell to inhibitory stimuli more prominent. In spontaneous tonic spiking solution there is an abrupt jump in frequency when the conductance of the CaL current becomes large enough. In hyperpolarization-induced bursting rhythms or PIR bursts, the CaL current allows the cell to generate substantially longer bursting responses. To summarize, the synergistic interaction of the CaT current and the CaL current enables an STN cell to respond to hyperpolarizing stimuli in a salient way, and this fact may implicate the important roles of the CaT current and the CaL current in the pathophysiology of the basal ganglia in disorders, such as in Parkinson's disease, noted for elevated burstiness of STN neurons.

**6.2. Bifurcation analysis and dynamical mechanisms of firing patterns.**

Bifurcation analysis allowed us to understand the effect of three currents (CaT, CaL, HCN) considered in this study on activity patterns of an STN cell under specific conditions. The availability of a current affects the structure of bifurcation diagram of fast subsystem, and the dynamics of slow variables with respect to the resulting bifurcation diagram yields an explanation of the mechanism underlying a specific activity pattern. More specifically, we found that the generation of various activity patterns depends on several factors: the relative position of the S-shaped curve of fixed points with respect to stable periodic orbits and $gtc$-nullcline, the place where the branches of stable periodic orbits terminate, and the existence of saddle-node bifurcation of periodic orbits (SNPO).

In this study we utilized four slow variables ($ghcnc$, $gtc$, $gcalc$, and $[Ca]$) for bifurcation analysis. Here, $[Ca]$ is the calcium concentration, $ghcnc = g_{HCN} * f$, $gtc = g_{CaT} * q$, $gcalc = g_{CaL} * d_1 * d_2$ where $f$, $q$, $d_{1,2}$ are gating variables and $g_{HCN}$, $g_{CaT}$, $g_{CaL}$ are maximal conductances for the HCN current, the CaT current, and the CaL current, respectively. In fact, these four slow variables are not sufficiently slow, so sometimes the projection of full model solution onto a bifurcation diagram shows some mismatch. In Figure 3D, for example, the projection of spiking solutions is not confined to the region inside two stable surfaces of stable orbits. If we make the four slow variables much slower, then we can resolve this mismatch, but the resulting activity patterns might not be physiologically realistic. Slow variables that were used in this study, however, were sufficiently slow so that we were able to obtain an insight into the underlying mechanism through bifurcation analysis.

Biophysicallly, the limit of very slow variables will lead to extremes in the firing patterns of a neuron (such as extremely long burst duration etc.). While one may argue that there is no sharp boundary between spiking and bursting activity, our results indicate that in the physiologically relevant regimes, the activity patterns exhibit specific bursting (which, if pushed to a limit in a mathematical consideration, leads to a coherent bifurcation structure).

**6.3 Concluding remarks**

As the only excitatory nucleus within the basal ganglia with strong pallidal and other inhibitory inputs, the fact that the STN is able to generate various bursting rhythms under hyperpolarization has an important implication in the pathophysiology of the basal ganglia. Synchronous beta oscillations within the basal ganglia is a hallmark of Parkinson's disease and has been associated with pathological symptoms related to movement [Brown, 2003; Hutchison, 2004; Kühn et al., 2004; Brown, 2007; Hammond et al., 2007; Mallet et al., 2008; Ray et al., 2008; Eusebio and Brown, 2009; Kühn et al., 2009; Park et al., 2010; Oswal et al., 2013; Stein and Bar-Gad, 2013; Ahn et al., 2015] Over the past decades, many theories have been developed with respect to the origin of the excessive beta rhythms within the basal ganglia (see Introduction). Two types of these theories focus on the role of the STN-GPe network and have been at the center of attention. In the first case, beta oscillations are generated in the cortex and the STN-GPe network in the basal ganglia has an ability to resonate or otherwise respond with oscillations at this frequency [see, for example, discussion in Stein and Bar-Gad, 2013]. In the second case, the STN-GPe network by itself has an ability to generate beta oscillations and plays an important role in maintaining the beta rhythms independently or via thalamus to the cortex connection [Bevan et al., 2002b; Mallet et al., 2008; Merrison-Hort and Borisyuk, 2013]. There are also studies suggesting that these situations are not mutually exclusive [Tachibana, 2011; Pavlides, 2015; Ahn et al., 2016]. Although whether the STN-GPe network generates the excessive beta rhythms in vivo in Parkinsonian brain or not is still uncertain, these two theories demonstrate the important role of the STN-GPe network in the excessive beta rhythmicity. In this context, the presently investigated dynamical mechanisms promoting the ability of an STN cell to generate bursting rhythms under either transient or sustained hyperpolarization may underlie excessively synchronous beta rhythms observed in Parkinsonian basal ganglia.

Finally, we would like to note the growing interest in the adaptive Deep Brain Stimulation (DBS) of the STN in Parkinson's disease. The development of effective control of beta-band activity may benefit from the availability of a relatively simple STN model like the one considered here, which captures the major dynamical characteristics of STN activity.

## Acknowledgements

This work was supported by NSF EHR 1700199 (CP), NSF DMS 1813819 (LLR), and IU-MSI STEM Initiative (CP and LLR).

## Data Availability

Data sharing is not applicable to this article as no new data were created or analyzed in this study.

## Captions

**Figure 1** Dependence of spontaneous tonic firing frequency on intrinsic parameters. We present the effects of (A) external constant input ($I_{app}$), (B) the CaT current ($g_{CaT}$), (C) the HCN current ($g_{HCN}$), and (D) the CaL current ($g_{CaL}$). In all four cases, the frequency increases as the current strength increases. In the panel (D), there is an abrupt increase in the frequency. Red plot in the same figure shows the result when the timescale of the dynamics of gating variable for the HCN current (*f*) was increased by 50%.

**Figure 2** (A) Bifurcation diagram of fast subsystem, which was projected onto $(gtc, V)$-space, with a bifurcation parameter $gtc$ when $g_{CaL} = 5$. Also, the $gtc$-nullcline (green) and the projection of the spontaneous tonic firing solution (blue) of the full model are shown for $g_{CaT} = 20$. (B) Bifurcation surface projected onto $(gtc, ghcnc, V)$-space with projection of a tonic firing solution trajectory when $g_{CaT} = 20$ (blue). (C) Two-parameter bifurcation diagram with RK (right knee) line (black slant line in the middle) and projection of four tonic firing solution trajectories (closed curves) when $g_{CaT} = 15$ (black), 20 (blue), 25 (red), and 40 (magenta). (D) Averaged $gtc$ ($\overline{gtc}$) over $gtc$ values within periodic orbit regime for $g_{CaT} = 20$ (blue), 30 (red), and 40 (magenta). Diagonal black line is $\overline{gtc} = gtc$ line. Here $ghcnc = 0.01$.

**Figure 3** (A - B) Bifurcation diagrams of fast subsystems in $(gtc, V)$-space with bifurcation parameter $gtc$ when $gcalc$ and $[Ca]$ are fixed. Here, $[Ca]$ is fixed at 0.05 in Fig 3A and at 0.18 in Fig 3B. In each figure, $gcalc$ = 2 (black), 6 (blue), and 10 (red). Green curve is $gtc$-nullcline for $g_{CaT} = 20$. (C) Two-parameter bifurcation diagrams with the projection of full model spiking solutions. Two vertical lines are RK lines when $[Ca]$ = 0.05 (black) and 0.18 (blue). Dotted lines that emanate from RK lines are SNPO lines. Short horizontal lines denote the projection of full model solutions. For smaller $g_{CaL}$ values, the projection of full model solutions lies on black RK lines (from bottom to top, $g_{CaL}$ = 5, 10, 15). For larger $g_{CaL}$ values, the full model solutions lie inside the regime of stable periodic orbits (from bottom to top, $g_{CaL}$ = 20 (black), 30 (blue), and 40 (red). (D) Bifurcation surface in ($gtc$, $gcalc$, $V$)-space when $[Ca]$ =0.18 with the projection of spiking solutions for $g_{CaL}$ = 20 (black), 30 (blue), and 40 (red), which were also shown in Fig. 3C. The S-shaped surface of fixed points (Σ) is shown in cyan. (E) Averaged $gtc$ ($\overline{gtc}$) values over various $[Ca]$ and $gcalc$ values: ($[Ca]$, $gcalc$) is (0.05, 2) for a black line, (0,05, 4) for a magenta line, (0.18, 8) for a brown line, and (0.18, 10) for a red line. Diagonal black line is $\overline{gtc} = gtc$ line.

**Figure 4** (A) Bursting rhythms for $g_{CaT}$ = 25 (black), 35 (blue), and 45 (red) with a fixed $g_{HCN}$ = 2. (B) Bursting rhythms for $g_{HCN}$ = 1 (black), 2 (blue), and 3 (red) with a fixed $g_{CaT}$ = 25. (C) Period, burst duration, and number of spikes within a burst in dependence on $g_{CaT}$ and $g_{HCN}$ conductances. (D) Bifurcation diagrams with bifurcation parameter $gtc$ for $ghcnc$ = 0.1 (black) and 0.2 (red). Green curve is the $gtc$-nullcline for $g_{CaT}$ = 25. (E) Bifurcation diagram in ($gtc$, $ghcnc$, $V$)-space with bifurcation parameters $gtc$ and $ghcnc$. Projection of bursting solution for $g_{CaT}$ = 25 and $g_{HCN}$ = 2 is shown in red. The S-shaped surface of fixed points is shown in cyan. (F) Two-parameter bifurcation diagram with projection of full model bursting solutions for $g_{CaT}$ = 25 (black), 35 (blue), and 45 (red) for $ghcnc$ = 2. (G) Two-parameter bifurcation diagram with projection of full model bursting solutions for $g_{HCN}$ = 1 (black), 2 (blue), and 3 (red) for $g_{CaT}$ = 25. In (F) and (G), black thin lines from left to right are HC line, RK line, and AH line.

**Figure 5** (A) Bursting rhythms for $g_{CaL}$ = 5 (black), 15 (blue), and 25 (red) with fixed $g_{CaT}$ = 25 and $g_{HCN}$ = 2. (B) Period (solid line), interburst interval (IBI) (dashed line), and burst duration (dotted line) as a function of $g_{CaL}$. (C) and (D) are period and burst duration plots over two-parameter space for $g_{CaL}$ = 5, 15, and 25 (from left to right). (E) Two exemplary bifurcation diagrams with bifurcation parameter $gtc$ with $[Ca]$ = 0.4 and $gcalc$ = 5 (dotted), and $[Ca]$ = 0.4 and $gcalc$ = 10 (solid). Green curve denotes $gtc$ -nullcline for $g_{CaT}$ = 25. (F) SNPO surface (left), AH surface (middle slant), and RK surface (right) in ($gtc$, $gcalc$, $[Ca]$)-space. The same figure also shows projections of three bursting solutions when $g_{CaL}$ = 5 (black), 15 (blue), and 25 (red).

**Figure 6** (A) An example of post-inhibitory rebound (PIR) burst. At $t$ = 500 msec, $I_{app}$ was changed from 0 to -20 for 500ms. (B-D) Bifurcation surfaces in ($gtc$, $ghcnc$, $V_m$)-space with the projection of full model PIR solution (blue) shown in (A). Green surface denotes $gtc$-nullsurface for $g_{CaT}$ = 20. (B) Spontaneous tonic spiking solution (from $t$ = 0 ms to $t$ = 500 ms). (C) The behavior of the trajectory under applied inhibitory current (from $t$ = 500 ms to $t$ = 1000 ms). The red curve on the lower surface of stable fixed points (Σ) is the intersection curve between this surface and $gtc$-nullsurface (Γ). (D) Activity pattern once the cell is released from inhibition. The cell jumps into the regime of stable periodic orbits. (E) Two-parameter bifurcation diagram with the projection of the PIR solution (blue) near RK line (black). When inhibition is turned on, the trajectory was pushed away from the RK line. (F) Activity patterns during and after the inhibition. Once the inhibition is turned on, RK line is shifted right upward (black dotted line on the upper right corner). The red line denotes Γ. Plot (E) is a magnification of the lower left corner of plot (F).

**Figure 7** The effect of magnitude and duration of inhibitory input on PIRs. (A) PIRs for $I_{app}$ = -5, -10, -20, and -30 from top to bottom. (B) Two-parameter diagram with the projection of solutions (thick closed curves) for $I_{app}$ = -5 (blue), -10 (red), -20 (magenta), and -30 (cyan). Black curve at lower left corner is the RK line for $I_{app} = 0$. Remaining thin curves are intersection curves of the lower surface of stable fixed points and $gtc$-nullsurface (Γs) introduced in the previous section for each value of $I_{app}$. Trajectory moves along the lower part of Σ to approach Γ under inhibition. These curves (Γs) use the same color code with the projection of full model solutions. (C) PIRs with four different input durations (100ms, 250ms, 500ms, and 750ms from top to bottom). (D) Two-parameter diagram with the projection of solutions (thick closed curves) when input durations are 100ms (blue), 250ms (red), 500ms (magenta), and 750ms (cyan). Black curve at lower left corner is the RK line and thin magenta curve at right upper corner is Γ. Termination of the inhibitory input is denoted by dot in each trajectory.

**Figure 8** Effect of CaT and CaL currents on PIRs for input duration 500ms and $I_{app} = -20$. (A - B) Effect of the CaT current on PIRs. (A) PIRs for $g_{CaT}$ = 15, 20, and 30 from top to bottom. (B) Two-parameter diagram with the projection of solutions (thick closed curves) when $g_{CaT}$ = 15 (blue), 20 (red), and 30 (magenta). Black curve at lower left corner is the RK line for $I_{app} = 0$. Remaining thin curves are intersection curves of the lower surface of stable fixed points and $gtc$-nullsurface (Γs) introduced in the previous section. Trajectory moves along the lower part of Σ to approach Γ under inhibition. These curves (Γs) use the same color code with the projection of full model solutions. (C - F) Effect of the CaL current on PIRs. (C) PIRs for $g_{CaL}$ = 5, 10, and 15 from top to bottom. (D) Two-parameter diagram with the projection of solutions and the corresponding intersection curves over the inhibitory input. $g_{CaL}$ = 5 (blue), 10 (red), and 15 (magenta). (E) Projection of trajectories in ($gtc$, $gcalc$, $[Ca]$) -space for the same values of $g_{CaL}$. SNPO/SNIC surface is also shown. (F) The same trajectories with SNPO/SNIC curve when $[Ca]$ = 0.05.

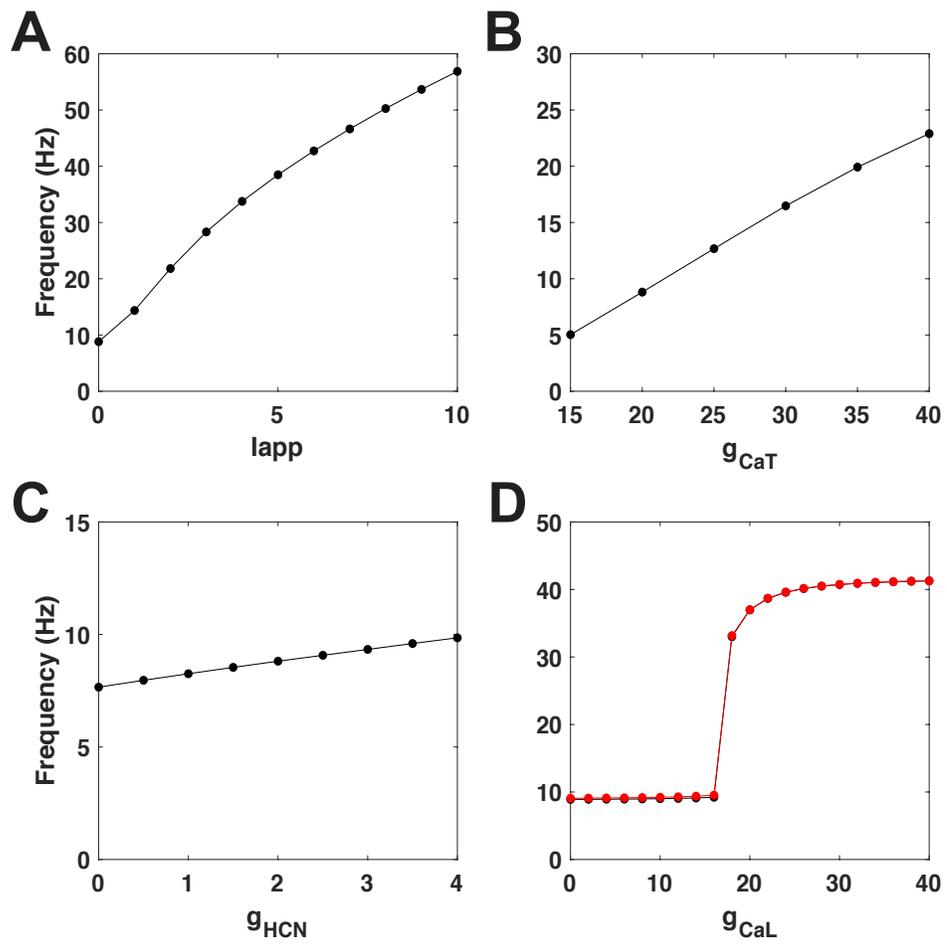

Figure 1

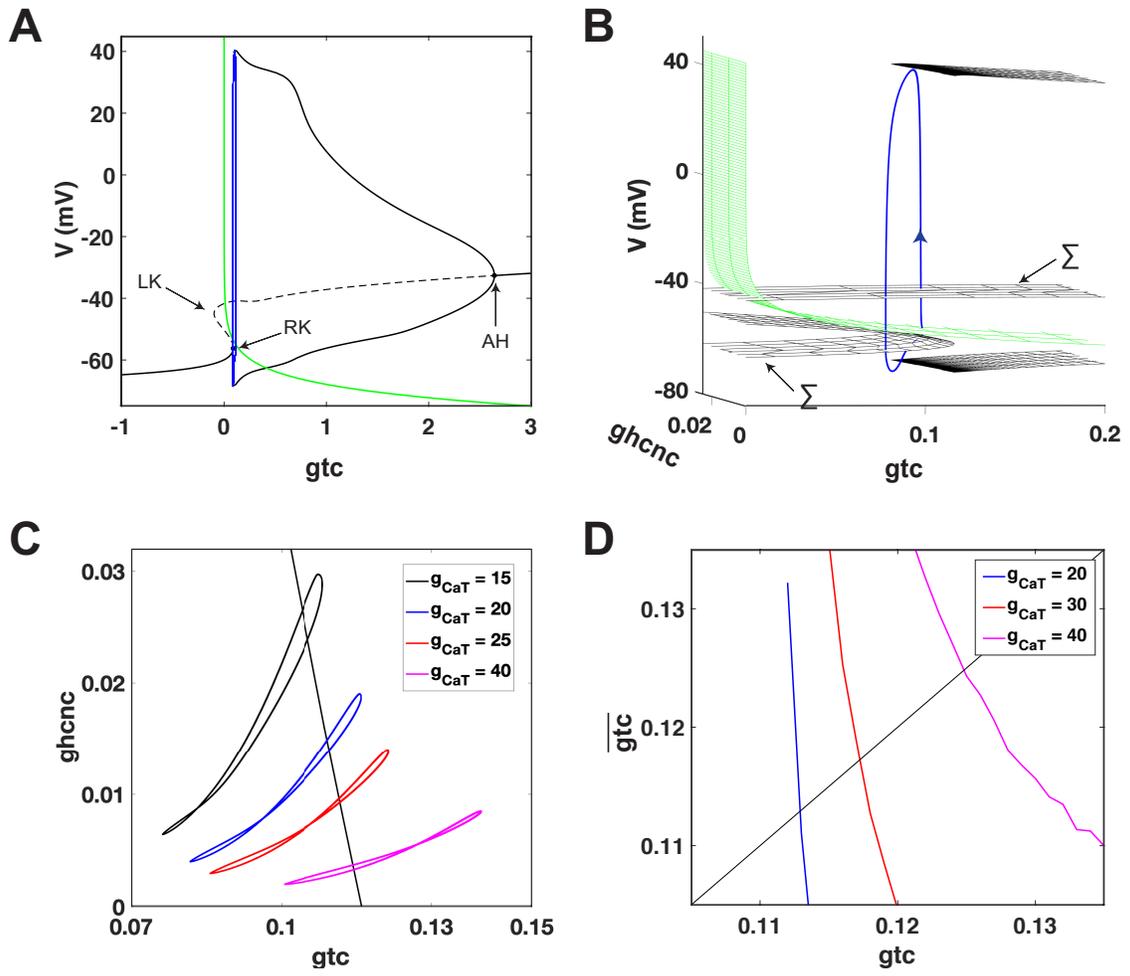

Figure 2

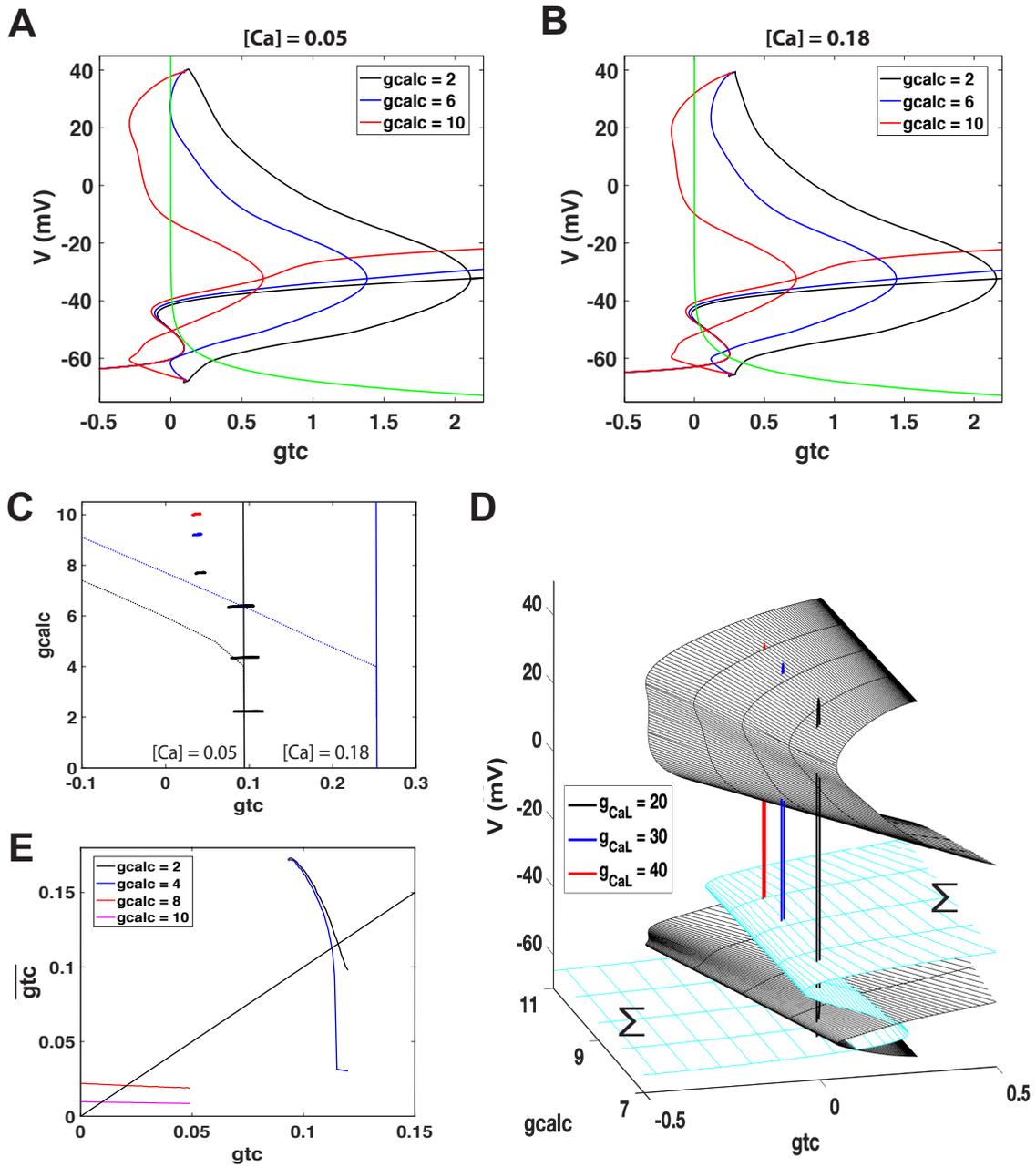

Figure 3

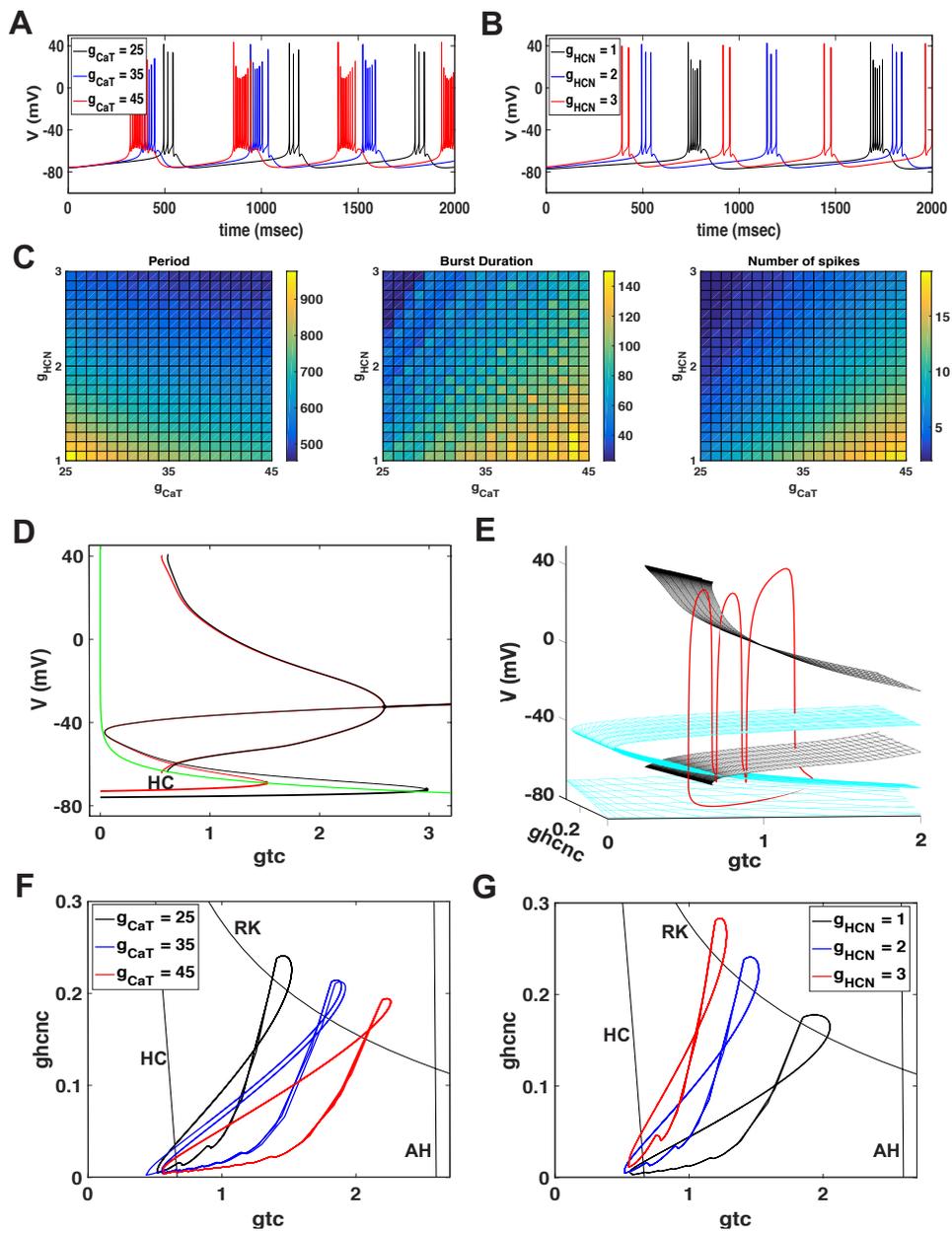

Figure 4

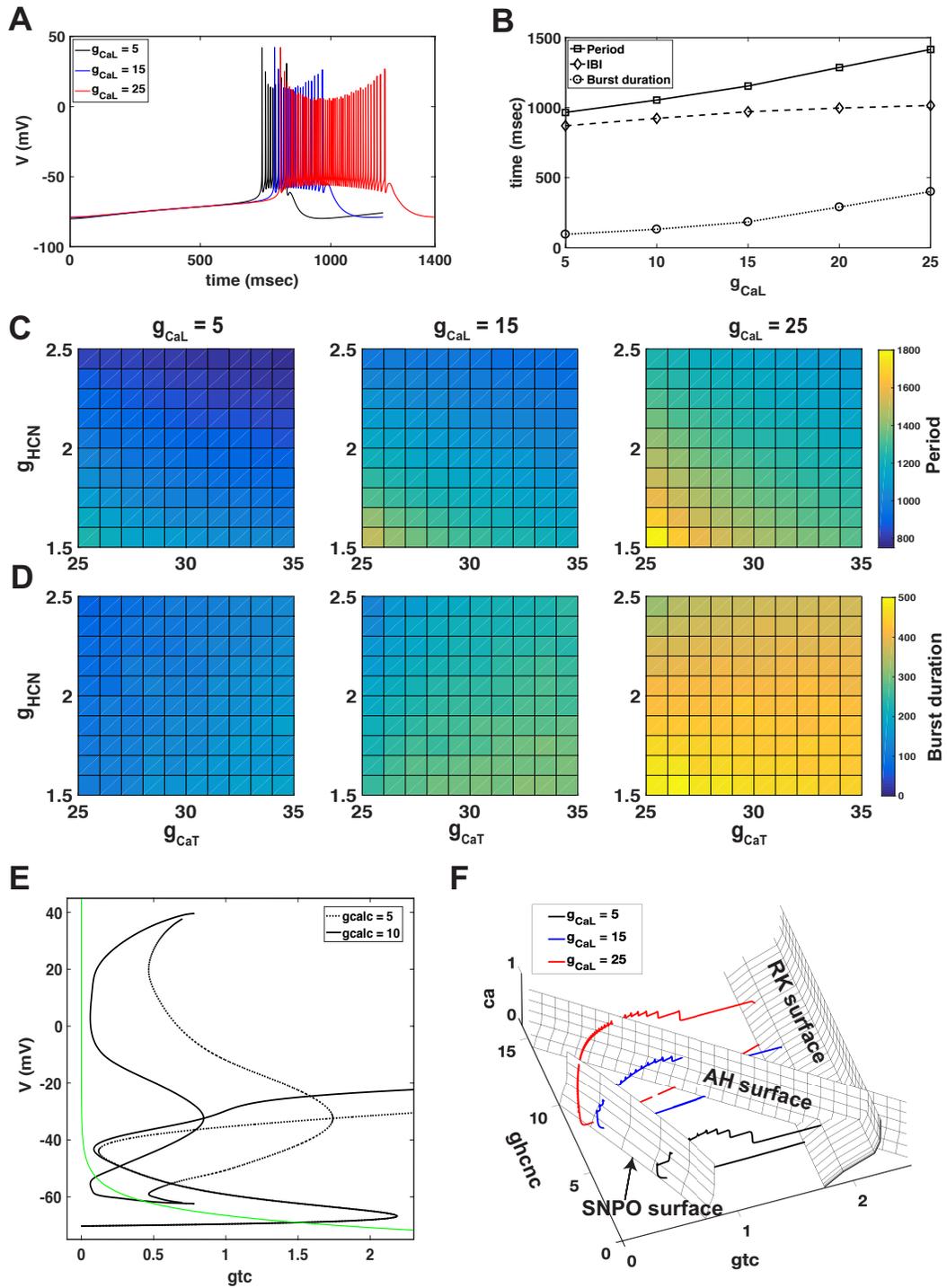

Figure 5

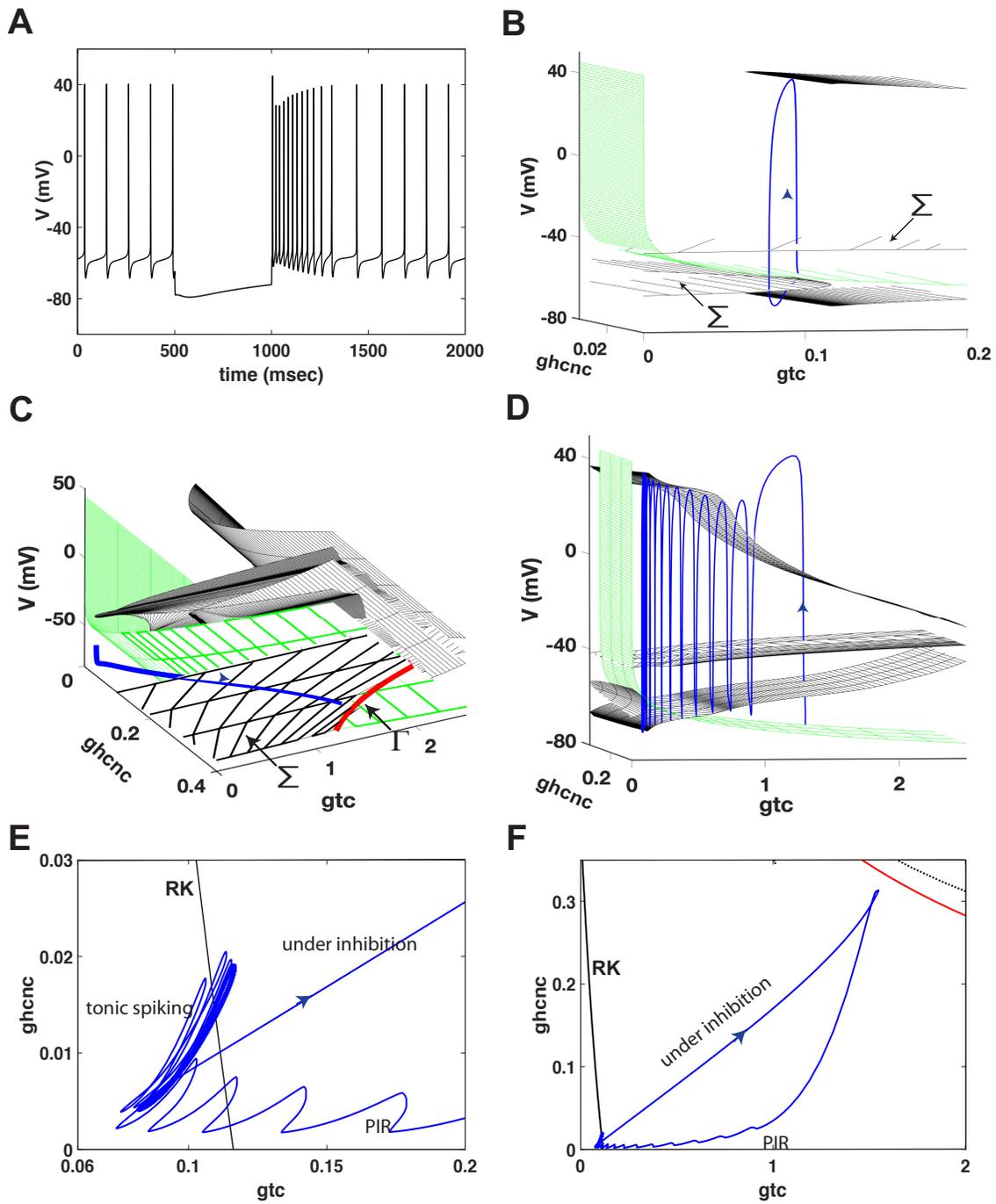

Figure 6

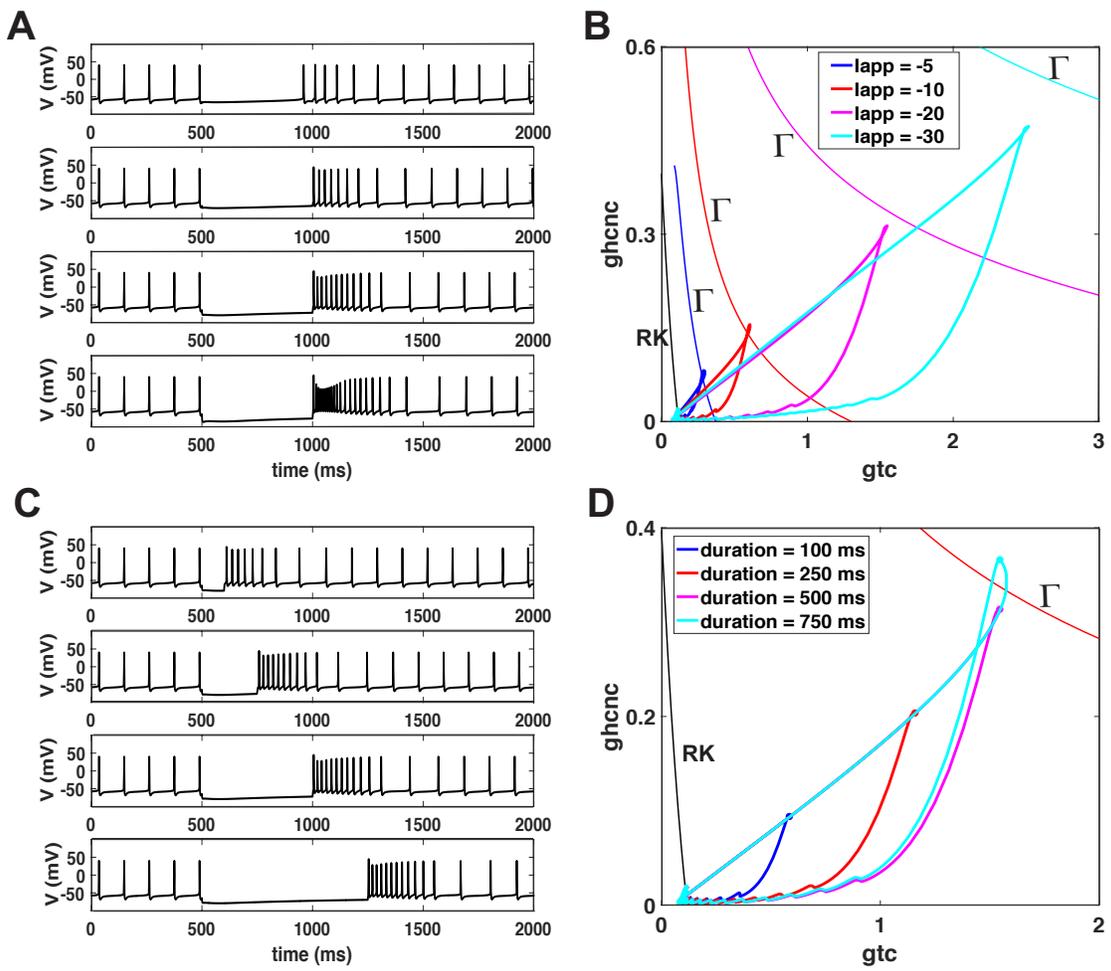

Figure 7

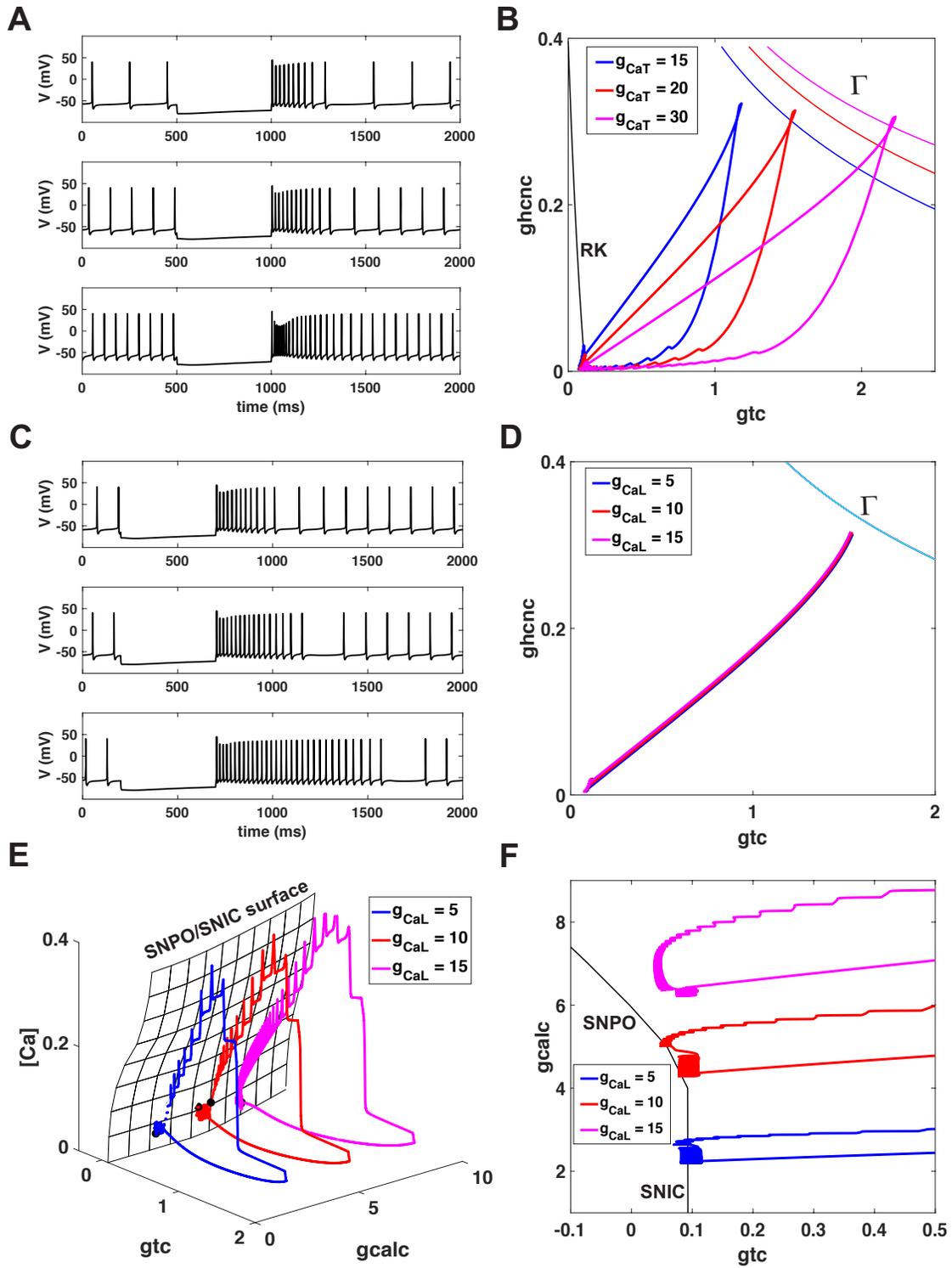

Figure 8